\documentclass{aa}  

\usepackage{graphicx}
\usepackage{txfonts}
\usepackage[colorlinks=true,linkcolor=blue,citecolor=blue,urlcolor=blue]{hyperref}
\usepackage{orcidlink}
\usepackage[flushleft]{threeparttable}
\usepackage{comment}
\usepackage{mathalfa}
\begin{document}

\title{General relativistic hydrodynamics of stellar tidal disruptions in Kerr spacetime: methods, validation, and first applications}

    \titlerunning{General relativistic hydrodynamics of stellar tidal disruptions}
    \authorrunning{Calderón \& Rosswog}

   \author{
        Diego Calderón
        \inst{1}
        \fnmsep
        \corrauth{calderon@mpa-garching.mpg.de}
        \fnmsep
        \thanks{Alexander von Humboldt Fellow}\orcidlink{0000-0002-9019-9951}
        \and
        Stephan Rosswog
        \inst{2}
        \fnmsep
        \inst{3}
        \email{stephan.rosswog@uni-hamburg.de}\orcidlink{0000-0002-3833-8520}
    }

   \institute{
        Max-Planck-Institut für Astrophysik, Karl-Schwarzschild-Straße 1, 85748 Garching, Germany
        \and Hamburger Sternwarte, Universit\"at Hamburg, Gojenbergsweg 112, 21029 Hamburg, Germany
        \and The Oskar Klein Centre, Department of Astronomy, AlbaNova, Stockholm University, SE-106 91 Stockholm, Sweden
    }

   \date{Received \today; accepted \today}

% \abstract{}{}{}{}{} 
% 5 {} token are mandatory
 
  \abstract
   {
   The disruption of a star by the tidal field of a super-massive black hole may provide insights into dormant and otherwise hard-to-study galactic nuclei. 
   The many physics facets and the wide spatial dynamic range of the problem make it extremely challenging to conduct global numerical studies. 
   State-of-the-art numerical tools have not converged yet on the importance of the strong relativistic effects in the disruption of stars by potentially spinning black holes.
   }
   {
   We present a specialised numerical tool to perform global hydrodynamic simulations of stellar tidal disruptions in curved spacetimes. 
   In addition, we quantify the role of impact strength and black hole spin onto the stellar structures and mass fallback rates.
   }
   {
   We adapted the numerical relativity code SPHINCS\_BSSN to perform General Relativistic Smoothed-Particle Hydrodynamics (GRSPH) simulations of stellar tidal disruptions in Kerr metric. 
   We coupled the code with a Newtonian self-gravity module that uses a recursive-coordinate-bisection tree, and we added the option to use an entropy evolution formulation to handle numerically challenging situations. 
   Besides describing the implementation and code validation, we present a set of 18 simulations of parabolic tidal disruptions of stellar polytropes to investigate the effect of impact strength and black hole spin.
   }
   {
   We demonstrate that SPHINCS is capable of performing GRSPH simulations, reproducing benchmark tests to machine precision. 
   Our stellar tidal disruption simulations show that the mass fallback rates agree with state-of-the-art general relativistic modelling. 
   Deep events (impact strength $\beta\gtrsim4$) result into stellar structures where self-gravity plays no role, the mass fallback rates peak at lower values, and rise-to-peak timescales decrease with impact strength. 
   Also, in these cases the black hole spin affects noticeable these quantities increasing (decrease) both fallback rate peak and rise-to-peak timescale for prograde (retrograde) spin.
   Last, fallback rates tend to decay with the characteristic (and expected) $t^{-5/3}$ on long timescales ($\gtrsim$1~yr).
   }
   {
   The results show that SPHINCS can simulate high-resolution stellar tidal disruptions in Kerr metric at a reasonable computational time. 
   This allows us to explore wide ranges of parameter space to model stellar tidal disruptions. 
   We discuss further guidelines and current numerical challenges. 
   }
   \maketitle
   \nolinenumbers
%
%-------------------------------------------------------------------
\section{Introduction}
    The disruption of a star due to the strong tidal field of a compact object generates a bright and characteristic flare known as tidal disruption event \citep[TDE;][]{hills1975}. 
    Such events occur in galactic centres, thus involving a super-massive black hole (SMBH) that typically is in a quiescent state and becomes active on day-to-week timescales \citep[see][for a review]{gezari2021}. 
    A TDE will take place once a star crosses  the so-called tidal radius $r_\text{t}$, i.e. the distance from the black hole where the tidal forces are stronger than the stellar self-gravity,
    \begin{equation}
        r_\text{t}\simeq R_\ast \left( \frac{M_\bullet}{m_\ast}\right)^{1/3}. 
    \end{equation}
    Here $M_\bullet$ is the black hole mass, and  $m_\ast$ and $R_\ast$ are the stellar mass and radius, respectively.
    As a result, roughly half of the stellar mass becomes unbound while the other half remains bound initially as a tidal stream, forms an accretion disc, and some of the material ends up being accreted \citep{hills1975,rees1988,alexander2005}. 
    This is the traditional scenario that assumes that the star is completely disrupted but there is evidence of variations in the case of partial disruption \citep[e.g.][]{phinney1989,guillochon2013,auchettl17,maguire20}, repeated partial disruptions \citep[e.g.][]{wevers23} as well as double disruptions of stellar binary systems which have been simulated \citep[e.g.][]{mandel2015,rosswog2020,mahapatra2026}.

    TDEs have the potential of informing us on the properties of quiescent SMBHs such as their mass and spin that otherwise would not be possible to measure directly. 
    Unfortunately, there are many unresolved puzzles about the physical mechanisms behind their observations.
    For instance, it is not clear how they are powered, if it is accretion itself \citep{metzger2016}, self-interaction of the tidal stream \citep{piran2015,jiang2016}, and/or secondary shocks during the circularisation phase \citep{bonnerot2020}. 
    Most TDEs have been detected in optical and ultraviolet (UV) wavelengths, which are very different to their expected peak in X-ray and UV. 
    This could imply that the light might be reprocessed in a quasi-spherical envelope that has been invoked as a result of the stellar disruption \citep[e.g.][]{metzger2016}. 
    In addition, there are some events that show X-ray emission but only a fraction of those have also been detected in optical range which have been {attributed to potential line of sight effects} of a unique phenomenon \citep{dai2018}. 
    To date, there are more than one hundred confirmed events \citep{vanvelzen2020,gezari2021,hammerstein2023,langis2026}. 
    However, this figure should see a dramatic increase thanks to the Legacy Survey of Space and Time (LSST) at the Vera C. Rubin Observatory \citep{ivezic2019,stone2020}. 
    Thus, it is a key priority to count with a solid theoretical understanding of the TDE phenomenology for interpreting the forthcoming data.

    Throughout the last decade the development of TDE numerical models has advanced tremendously. 
    The first global hydrodynamic simulations of stellar tidal  disruptions were developed by \cite{evans1989}. 
    Since then, numerical hydrodynamic modelling has progressed making use of both Eulerian and Lagrangian approaches, and even of hybrid techniques. 
    State-of-the-art works making use of the Eulerian approach have opted for developing alternatives techniques to maximise the finite-volume resolution, namely simulating the star in a box moving along its orbit and then remapping it into a larger domain to follow the longer-term evolution \citep[e.g.][]{guillochon2013,ryu2023,abolmasov2026}.
    However, in general Lagrangian approaches are chosen for conducting global simulations of stellar tidal disruptions, as they have the advantage of not being confined with a limited spatial domain. 
    In this context, employed codes include  the moving-mesh code Arepo \citep{springel2010}, the smoothed-particle hydrodynamics (SPH) code Gadget \citep{springel2021}, the mesh-free code GIZMO \citep{hopkins2015}, the SPH code MAGMA/MAGMA2 \citep{rosswog2007,rosswog2020}, the SPH code Phantom \citep{price2018}, and the moving-mesh code RICH \citep{yalinewich2015}. 
    This wide variety of options has led to the addition of more complex physical processes in the models. 
    For instance, the consideration of realistic stellar structures \citep{goicovic2019,jankovic2023},  nuclear burning \citep{rosswog2009,tanikawa2017}, magnetic fields \citep{bonnerot2017,pacuraru2026}, relativistic dynamics \citep{tejeda2013,cheng2014,tejeda2017,gafton2019}, radiation treatment \citep{bonnerot2021,steinberg2024}, and hydrodynamics in curved spacetimes \citep{liptai2019b,jankovic2023,price2024}. 
    Yet the availability of codes to perform hydrodynamic simulations of stellar tidal disruption in Kerr metric is limited. 
    The most widely used tools are the codes Phantom \citep{price2018,liptai2019}, GIZMO \citep{hopkins2015,lupi2023}, and AREPO \citep{springel2010,pakmor2016}. 
    Thus, there is a need to provide a benchmark of stellar disruptions due to the tidal forces of a central object in Kerr metric in relatively high-resolution. 
    So that it is possible to build on it adding complexity progressively to provide a solid base for a converged understanding of the general relativistic effects on TDEs.

    The development of the general relativistic smoothed-particle hydrodynamic (GRSPH) code SPHINCS\_BSSN has provided the first Lagrangian numerical tool for modelling stellar mergers in curved spacetimes evolving fully self-consistently both the fluid dynamics and the spacetime \citep{rosswog2021,rosswog2023}. 
    In this work, we have expanded the capabilities of the code, adapting it to simulate efficiently tidal disruptions of stars by a central object in time-independent spacetimes. 
    Specifically, we have developed a module for using the Kerr metric to determine the curvature of the spacetime, and coupled it with a self-gravity module for setting up stellar profiles in hydrostatic equilibrium. 
    In addition, we added the option to use the entropy formulation of the GRSPH equations, as this aids numerically with the highly relativistic motion of the tidally disrupted star. 
    Finally, we present the first application of this tool: a Solar-type star on a parabolic orbit being tidally disrupted by a SMBH. 
    We study the role of the impact strength $\beta$ of the event and the black hole spin, where
    \begin{equation}
        \beta=\frac{r_\text{t}}{r_\text{p}} 
        \label{eq:beta}
    \end{equation}
    is a measure of the impact strength, and $r_\text{p}$ is the pericentre distance. 
    Overall, the results show that deeper events ($\beta\gtrsim4$) show a lower fallback rate peak value, faster rising to peak but longer duration. 
    In general, a prograde (retrograde) rapidly spinning SMBH causes a slight increase (decrease) of the peak of the mass fallback rates. 
    But the spin effect on the peak time or rise-to-peak timescale is negligible for shallow events ($\beta\lesssim4$). 
    Only in deep events a prograde (retrograde) spin increases (decreases) these timescales due to the gravitational potential causing a significantly wider (narrower) energy spread of the stellar stream. 
    Nevertheless, the late-time evolution ends up following the decay $\propto t^{-5/3}$ in all cases, as expected for full stellar disruptions in Newtonian gravity. 

    This work is presented as follows: in Section~\ref{sec:method} we describe the numerical formalism of SPHINCS together with its new implementations: the use of the (time-independent) Kerr metric, the self-gravity module, and their coupling. 
    Section~\ref{sec:validation} contains the validation tests of the implementation and coupling of the modules.
    In Section~\ref{sec:application}, we present the application of the code to model stellar tidal disruptions in Kerr metric.
    The results of these numerical simulations are presented in Section~\ref{sec:results}.
    Then, we compare the results with previous works from the literature and discuss current numerical challenges in Section~\ref{sec:disc}. 
    Finally, we present the conclusions and future applications in Section~\ref{sec:conclusions}.
\section{Numerical simulations}
\label{sec:method}
    Our simulations were conducted using an adapted version of the Lagrangian Numerical Relativity code SPHINCS\_BSSN\footnote{SPHINCS stands for ``Smoothed Particle Hydrodynamics In Curved Spacetime", where ``\_BSSN" indicates that the dynamical spacetime evolution uses the Baumgarte-Shapiro-Shibata-Nakamura (BSSN) formulation \citep[see e.g.][]{Baumgarte2010}. Since here the spacetime is not evolved, we simply refer to this code version as SPHINCS.}, which solves the Lagrangian hydrodynamic equations in a curved spacetime \citep{rosswog2021,rosswog2023}. 
    For a recent detailed review of SPH in general and on its relativistic version, see \cite{rosswog2026}.
    Here, we give a brief summary of the numerical approach of SPHINCS and its time-independent metric module. 
    \subsection{SPHINCS formalism}
        We perform the simulations in a fixed computing frame with coordinates $(t,x,y,z)$ and with  metric $g_{\mu\nu}$. 
        Greek indices take values $(0,1,2,3)$ and Latin indices run over $(1,2,3)$. 
        Unless stated otherwise, we use $G=c=1$. 
        We start introducing the generalised Lorentz factor
        \begin{equation}
            \Theta=\frac{1}{\sqrt{-g_{\mu\nu}v^{\mu}v^{\nu}}},
        \end{equation}
        where $v^{\alpha}$ are the coordinate velocities, i.e.
        \begin{equation}
            v^{\alpha}=\frac{dx^{\alpha}}{dt}.
        \end{equation}
        These are related to the four-velocities $U^{\alpha}$ using the normalisation $U_{\alpha}U^{\alpha}=-1$,
        \begin{equation}
            v^{\alpha}=\frac{dx^{\alpha}}{dt}=\frac{U^{\alpha}}{\Theta}=\frac{U^{\alpha}}{U^0}.
            \label{eq:coord_vel}
        \end{equation}
        In SPHINCS, the fluid is described using the mass density in the computing frame $\rho^*$ that is related to the mass density measured in the local fluid rest frame $\rho$ through
        \begin{equation}
            \rho^* = \sqrt{-g}\Theta~\rho,
            \label{eq:rhostar}
        \end{equation}
        where $g$ is the determinant of the metric.
        The mass of each SPH particle $m$ is constant, so that the total mass is conserved. 
        At every timestep, the computing frame mass density at the position of an arbitrary particle $a$ is calculated via a weighted sum
        \begin{equation}
            \rho^*_a=\sum_b~m_b W(|\vec{r}_a-\vec{r}_b|,h_a),
            \label{eq:density}
        \end{equation}
        where $b$ is the sum index, $\vec{r}=(x,y,z)$ is the position vector, and $h$ is the smoothing length that defines the size of the SPH kernel $W$.
        For the fluid momentum and energy variables we use the canonical momentum $S_i$ and the canonical energy $e$, respectively;
        \begin{eqnarray}
            \left(S_i\right)_a
            &=&
            \left(\Theta\mathcal{E}v_i\right)_a,
            \\
            e_a
            &=&
            \left(S_iv^i+\frac{1+u}{\Theta}\right)_a=\left(\Theta\mathcal{E}v_iv^i+\frac{1+u}{\Theta}\right)_a.
        \end{eqnarray}
        These are derived from the Lagrangian of an ideal fluid \citep[see][for a step-by-step derivation]{rosswog2009}.
        Here we have introduced the specific relativistic enthalpy $\mathcal{E} = 1+u+P/\rho$, with $u$ being the specific internal energy, and $P$ the thermal pressure of the fluid.
        The momentum and energy variables are evolved in time through the equations
        \begin{eqnarray}
            \label{eq:momentum}
            \frac{d\left(S_i\right)_a}{dt}
            &=&
            -\sum_b~m_b\left\{\frac{P_a+Q_a}{\rho_a^{*2}}D^a_i+\frac{P_b+Q_b}{\rho_b^{*2}}D_i^b\right\}
            \nonumber
            \\
            &&
            +\left(\frac{\sqrt{-g}}{2\rho^*}T^{\mu\nu}\frac{\partial g_{\mu\nu}}{\partial x^i}\right)_a,
            \nonumber
            \\
            &&
            \\
            \label{eq:energy}
            \frac{de_a}{dt}
            &=&
            -\sum_b~m_b\left\{\frac{P_a+Q_a}{\rho_a^{*2}}v_b^iD^a_i+\frac{P_b+Q_b}{\rho_b^{*2}}v_a^iD_i^b\right\}
            \nonumber
            \\
            &&
            -\left(\frac{\sqrt{-g}}{2\rho^*}T^{\mu\nu}\frac{\partial g_{\mu\nu}}{\partial t}\right)_a+\Pi^a_\text{cond},
            \nonumber\\ 
            &&
        \end{eqnarray}
        where $Q_a$ and $Q_b$ are the viscous pressures due to shock dissipation, $\Pi^a_\text{cond}$ represents the artificial conductivity, and $T^{\mu\nu}=\left\{\rho\left[1+u\right]+P\right\}U^{\mu}U^{\nu}+Pg^{\mu\nu}$ is the energy-momentum tensor
        of an ideal fluid. 
        In addition, we have abbreviated
        \begin{eqnarray}
            D_i^a
            &=& 
            \sqrt{-g_a}\frac{\partial W_{ab}(h_a)}{\partial x^i_a},~{\rm and}
            \\
            D_i^b
            &=&
            \sqrt{-g_b}\frac{\partial W_{ab}(h_b)}{\partial x^i_b}.
        \end{eqnarray}
        Here, the kernel notation is $W_{ab}(h_k)=W(|\vec{r}_a-\vec{r}_b|/h_k)$.         
        For the viscous pressures we have followed mostly \cite{liptai2019}:
        \begin{eqnarray}
            Q_a
            &=&
            -\frac{1}{2}\alpha_\text{AV}\rho^*_av_\text{s,a}\mathcal{E}_a\left(\Gamma^*_aV_a^*-\Gamma^*_bV_b^*\right)
            \\
            Q_b
            &=&
            -\frac{1}{2}\alpha_\text{AV}\rho^*_av_\text{s,b}\mathcal{E}_b\left(\Gamma^*_aV_a^*-\Gamma^*_bV_b^*\right),\label{eq:Qvis}
        \end{eqnarray}
        where $V^*$ is the velocity of an Eulerian observer along the line connecting the particles $a$ and $b$, i.e. $V^*_a=\eta_{ij}\hat{e}^j_{ab}V^i_a$; and $\Gamma^*_a=1/\sqrt{1-V^{*2}_a}$. 
        The Eulerian and coordinate velocities are related through
        \begin{equation}
            V^i=\frac{v^i+\beta^i}{\alpha,}
        \end{equation}
        where $\alpha$ and $\beta^i$ are the lapse function and the shift vector, respectively.
        The signal speed is defined as
        \begin{equation}
            v_{\text{s,}a}=\frac{c_{\text{s,}a}+|V_{ab}^*|}{1+c_{\text{s,}a}|V_{ab}^*|},
        \end{equation}
        where the relativistic sound speed is $c_\text{s}=\sqrt{(\Gamma-1)(\mathcal{E}-1)/\mathcal{E}}$, and
        \begin{equation}
            V^*_{ab}=\frac{V^*_a-V^*_b}{1-V^*_aV^*_b}.
        \end{equation}
        The artificial conductivity is given by
        \begin{equation}
            \Pi^a_\text{cond}=\frac{\alpha_u}{2}\sum_bm_b\xi^u_{ab}\left(\frac{\alpha_au_a}{\Gamma_a}-\frac{\alpha_bu_b}{\Gamma_b}\right)\left\{\frac{v_{\text{s},a}^uD^a_i}{\rho_a^*}+\frac{v_{\text{s},b}^uD^b_i}{\rho_b^*}\right\}\hat{e}^i_{ab},
            \label{eq:AC}
        \end{equation}
        where $\alpha_a$ and $\alpha_b$ are the lapse functions at the particle positions and $\Gamma=(1-V_iV^i)^{-1/2}$. 
        Apart from the limiter $\xi^u_{ab}$ (see below), the conductivity term is the same as in \cite{liptai2019}.
        The conductivity signal velocity is
        \begin{equation}
            v_\text{s}^u=\min\left\{1,\sqrt{\frac{2|P_a-P_b|}{\mathcal{E}_a\rho_a+\mathcal{E}_b\rho_b}}\right\},
        \end{equation}
        and we use the prefactor $\alpha_u=0.3$.  We also use the conductivity limiter introduced in \cite{rosswog2021},
        \begin{equation}
            \xi^u_{ab}=\frac{T_{u,ab}}{T_{u,ab}+0.01}, \quad {\rm with} \quad T_{u,ab}=\frac{h_{ab}}{u_{ab}}\left|\left(\nabla u\right)_a-\left(\nabla u\right)_b\right|
        \end{equation}
        where $u_{ab}=(u_a+u_b)/2$ and $h_{ab}=(h_a+h_b)/2$. When the dimensionless quantity $\xi^u_{ab}$ is large, the conductivity is switched on, otherwise it is suppressed. 
        When simulating self-gravitating fluids, hydrostatic equilibrium is more easily achievable when $v^u_s=\left|V^*_{ab}\right|$ is used as conductivity signal speed \citep{price2018,liptai2019}. 
        Notice that unlike in \cite{liptai2019}, we use a linear reconstruction of the jumping quantities, i.e. differences between quantities a particle $a$ and particle $b$ in equations~(\ref{eq:Qvis}) and (\ref{eq:AC}), which suppresses dissipation where it is not needed. 
        For more details we refer the interested reader to Section~2.2.3 in \cite{rosswog2021}.
        
        The set of the equations~(\ref{eq:density}),~(\ref{eq:momentum}), and~(\ref{eq:energy}) describes the evolution of the fluid variables $(\rho^*,S_i,e)$. 
        To relate them to the physical properties $(\rho,v^{\alpha},u)$ we still need an equation of state. 
        A simple but useful choice is a polytropic equation of state $P=(\Gamma-1)\rho u$, where $\Gamma$ is the adiabatic exponent. 
        It is important to remark that the fluid quantities are measured in their local rest frame.  
        From the large variety of kernel functions implemented in SPHINCS we choose the C6-smooth Wendland kernel \citep{wendland1995} with exactly $300$ contributing neighbours in the support of each particle's kernel which is a very good choice based on precision experiments \citep{rosswog2015a,rosswog2026}.

        For certain problems, the use of the conservative quantities $(\rho^*,S_i,e)$ can lead to problems since the positivity of the (specific) internal energy (and therefore pressure) is not always guaranteed. 
        Based on this, we have included the option of evolving the (pseudo-)entropy quantity $K$ instead of the canonical energy $e$ \citep{springel2002}. 
        To do so, we have implemented the approach presented by \cite{liptai2019} that evolves the conservative quantities $(\rho^*,S_i,K)$, where the (pseudo-)entropy is defined by the polytropic relation $P=K\rho^{\Gamma}$. 
        The general relativistic entropy equation can be derived from the second law of thermodynamics \citep[see Section 5 in][]{liptai2019}, and is given by
        \begin{equation}
            \frac{dK_a}{dt}=\frac{\Theta_a K_a}{u_a}\left\{\Pi^a_\text{cond}+\sum_bm_b\frac{Q_a(v_a^i-v_b^i)}{\rho^{*2}_a}D_i^a\right\}.
        \end{equation}
        Note that entropy changes are entirely due to dissipative terms.  
        When using the entropy (instead of canonical energy) as the conserved quantity it was necessary to adapt the recovery routines to compute the primitive variables from the conserved variables after every Runge-Kutta sub-step.
        In this case, the recovery is done via the enthalpy (instead of pressure) following the method developed by \cite{tejeda2012}. 
        The exact implementation is documented in Appendices~\ref{app:recovery} and~\ref{app:enthalpy}.
    \subsection{SPHINCS with time-independent metric}
        In this work, we use the Kerr metric \citep{kerr1963} expressed in Cartesian Kerr-Schild (CKS) coordinates
        with line element
        \begin{eqnarray}
            ds^2
            &=&
            -dt^2+dx^2+dy^2+dz^2\nonumber\\
            &&
            +\frac{2Mr}{\rho^2_\text{met}}\left[\frac{r(xdx+ydy)-a(xdy-ydx)}{r^2+a^2}+\frac{zdz}{r}+dt\right]^2,\nonumber\\
        \end{eqnarray}
        where $M$ corresponds to the central object mass, $a$ to its spin parameter\footnote{Not to be confused with the dimensionless spin parameter $\chi=a/M$ that satisfies $\left|\chi\right|\leq1$.}, and\footnote{The subscript ``met" in $\rho_\text{met}$ is not to confuse it with the density in the local rest frame $\rho$.}
        \begin{equation}
            \rho^2_\text{met}=r^2+\frac{a^2z^2}{r^2}=\sqrt{(x^2+y^2+z^2-a^2)^2+4a^2z^2}.
            \label{eq:rho_metric}
        \end{equation}
        The evolution equations (\ref{eq:momentum}) and (\ref{eq:energy}) require derivatives of the metric for their spacetime contributions, but since we use a time-independent Kerr metric, we of course only need the spatial derivatives which we explicitly provide in Appendices~\ref{app:metric} and~\ref{app:dmetric}.
    \subsection{Self-gravity}
        For a SMBH and a solar-type star, the stellar self-gravity is only
        a very small perturbation of the background metric, $|h_{\mu\nu}|\ll|g_{\mu\nu}|$.  
        To leading order, the metric perturbation is \citep{poisson2014}
        \begin{eqnarray}
            h_{00}
            &=&
            -2\Phi,
            \\
            h_{ij}
            &=&
            2\Phi \, \delta_{ij},
        \end{eqnarray}
        where $\Phi$ is the Newtonian gravitational potential, and $\delta_{ij}$ is the Kronecker delta.
        Notice that $|h_{\mu\nu}|\ll|g_{\mu\nu}|$ does not imply that the gradient of the perturbation $h_{\mu\nu}$ negligible. 
        Then, the contribution of the self-gravity to the momentum equation is
        \begin{equation}
            \left(\frac{d\left(S_i\right)_a}{dt}\right)_\text{sg}
            =
            \left\{\frac{1}{2}\left[\Theta\left(1+u\right)+\frac{P}{\rho}\right]v^{\mu}v^{\nu}\frac{\partial h_{\mu\nu}}{\partial x^i}\right\}_a. 
        \end{equation}
        Replacing the expression for the perturbation
        \begin{equation}
            \left(\frac{d\left(S_i\right)_a}{dt}\right)_\text{sg}
            =
            \left[\Theta_a\left(1+u_a\right)+\frac{P_a}{\rho_a}\right]\left[(v^jv_{j})_a-1\right]\frac{\partial\Phi_a}{\partial x^i}.
        \end{equation}
        The gravitational potential $\Phi$ is calculated using the approach from the code MAGMA2 \citep[see Section~2.4 from][]{rosswog2020}. 
        For a fast calculation of gravitational forces and potentials, we use the very fast ``Recursive Coordinate Bisection" (RCB) tree \citep{gafton2011}\footnote{To avoid misunderstandings, we want to stress that this is (for speed reasons) a {\em non-recursive} tree, it is only the coordinate bisection that is recursive.}.
    \subsection{Integration scheme}
        The set of equations is integrated numerically through an optimal 3rd order total variation diminishing Runge–Kutta scheme \citep{gottlieb1998}. 
        That is, the numerical variables are collected into the vector $\vec{Y}=[x^i,S_i,e]$ (or $[x^i,S_i,K]$) that is evolved as follows
        \begin{eqnarray}
            \vec{Y}^{(1)}
            &=&
            \vec{Y}^{n}+\Delta tL(\vec{Y}^n)
            \\
            \vec{Y}^{(2)}
            &=&
            \frac{1}{4}\left[3\vec{Y}^n+\vec{Y}^{(1)}+\Delta tL(\vec{Y}^{(1)})\right]
            \\
            \vec{Y}^{n+1}
            &=&
            \frac{1}{3}\left[\vec{Y}^n+2\vec{Y}^{(2)}+2\Delta tL(\vec{Y}^{(2)})\right],
        \end{eqnarray}
        where the function $L=L(\vec{Y})$ corresponds to the time derivatives of the vector $\vec{Y}$, and $\Delta t$ to the discrete timestep. 
    \subsection{Timestep constrain}
    \label{sec:dt}
        To ensure the stable and accurate  time integration the timestep is constrained through a number of criteria; 
        \begin{eqnarray}
            \Delta t_a^\text{hyd}
            &=&
            \min\left\{\frac{h_a}{v_\text{s,a}},\sqrt{h_a\left|\left(\frac{d\mathbf{S}_a}{dt}\right)_\text{hyd}\right|^{-1}}\right\},
            \\
            \Delta t_a^\text{met}
            &=&
            \sqrt{h_a\left|\left(\frac{d\mathbf{S}_a}{dt}\right)_\text{met}\right|^{-1}},
            \\
            \Delta t_a^\text{div}
            &=&
            0.05\left|\partial_i v^i\right|_a^{-1},
            \\
            \Delta t_a^\text{bh}
            &=&
            0.01\sqrt{r_a\left|\left(\frac{d\mathbf{S}_a}{dt}\right)_\text{met}\right|^{-1}},
        \end{eqnarray}
        where the subscripts ``hyd" and ``met" refer to the hydrodynamic and metric contributions to the momentum equation, respectively (see equation~\ref{eq:momentum}). 
        The timestep constraints include the Courant condition \citep{press1992}, various criteria triggering of the particle acceleration ($\propto |d\mathbf{S}/dt|$), and a criterion sensitive to compression/expansion (involving $\partial_i v^i$).
        The resulting global timestep is set to the minimum among these constraints across all the particles multiplied by the Courant-type pre-factor $C=0.2$, i.e. 
        \begin{equation}
            \Delta t= C\min\{\Delta t^\text{hyd}_a,\Delta t^\text{met}_a,\Delta t^\text{div}_a,\Delta t^\text{bh}_a\}.
        \end{equation}
        \begin{figure*}
            \centering
            \includegraphics[width=0.85\linewidth]{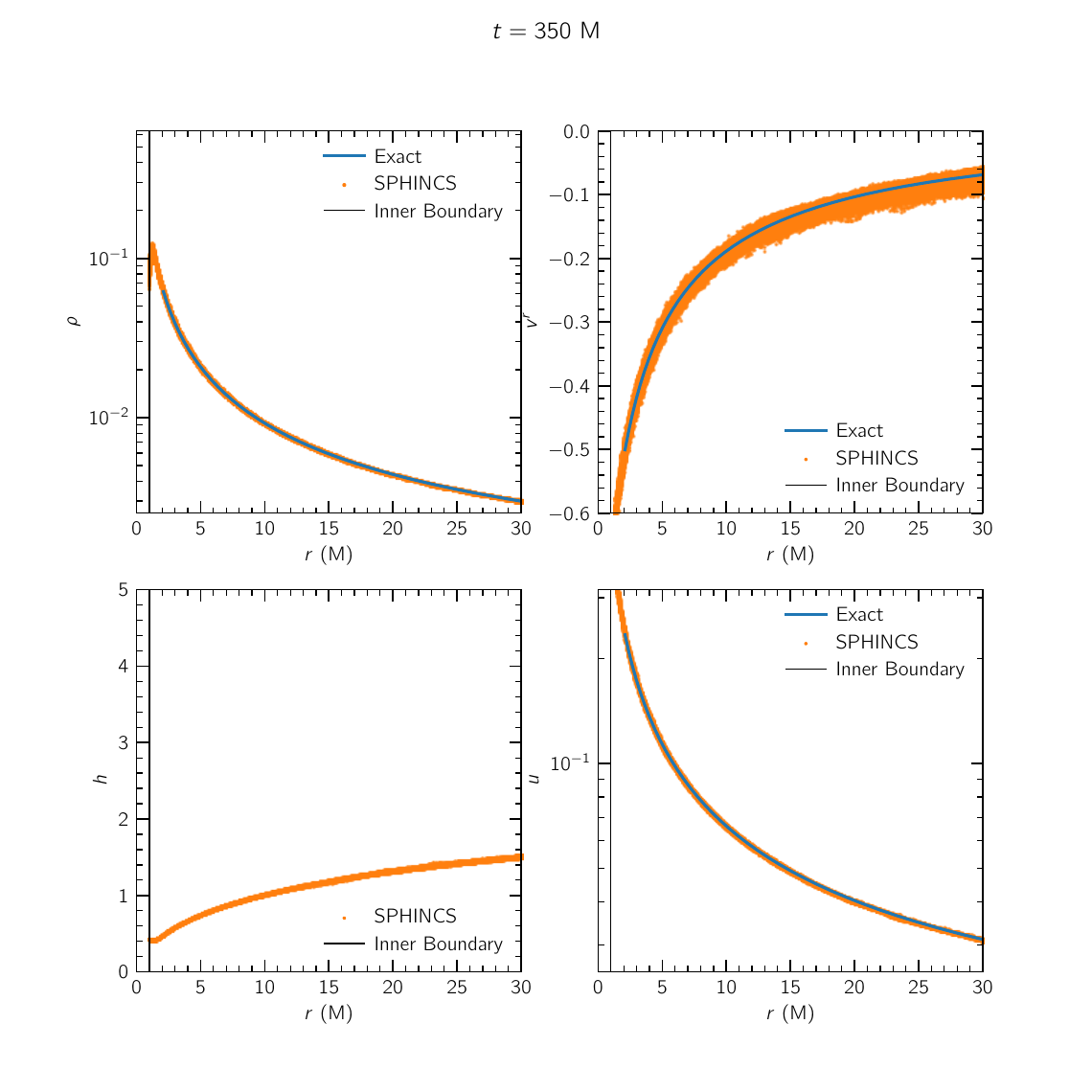}
            \caption{
            Spherical accretion test in strong gravity in Kerr metric with $M=1$ and $a=0$. 
            Top left-, top right-, bottom left-, and bottom right-hand side panels show the radial profiles of mass density in the local rest frame, radial velocity, smoothing length, and specific internal energy, respectively. 
            The blue lines represent the exact solution of the spherical accretion and the orange dots show the results from the numerical simulation.}
            \label{fig:bondi}
        \end{figure*}
\section{Validation Tests}
\label{sec:validation}
    To ensure the correct implementation of the time-independent metric module in SPHINCS, we have developed a set of tests that were motivated by  codes such as Phantom \citep{price2018,liptai2019} and GIZMO \citep{hopkins2015,lupi2023}. 
    We performed several tests of particle dynamics in Kerr metric such as circular stable orbits, radial geodesics, pericentre shifts, and epicyclic motion. 
    In Appendix~\ref{app:tests}, we summarise these benchmark tests which were all passed with very good accuracy.
    To ensure the correct coupling of the module with the hydrodynamics we simulated spherical accretion in strong gravity regime in Kerr metric that we present as follows.
    \subsection{Spherical accretion}
        The problem of spherical accretion under the influence of a strong gravitational field has been studied and described analytically \citep{michel1972,hawley1984}. 
        The solution is a generalisation of the Bondi accretion problem in steady state. 
        To describe the solution, let us consider a spherically symmetric function $T=T(r)$ defined as $T=P/\rho=(\Gamma-1)u$. 
        Then, in steady state the properties of the fluid infalling towards a central object with $M=1$ in Schwarzschild metric are
        \begin{eqnarray}
            U^r(r)
            &=&
            \frac{C_1}{r^2T^q(r)},
            \\
            \rho(r)
            &=&
            K_0T^q(r),
            \\
            u(r)
            &=&
            qT(r),
        \end{eqnarray}
        where $q=1/(\Gamma-1)$ and the function $T$ can be obtained solving the implicit equation
        \begin{equation}
            C_2=\left[1+(q+1)T(r)\right]^2\left\{1-\frac{2M}{r}+\left[U^r(r)\right]^2\right\},
        \end{equation}
        where $C_1$ and $C_2$ are constant values. 
        Thus, the solution can be obtained once both constants are calculated. 
        This can be done through the analysis of values of the fields at the critical point $r_\text{c}$. 
        Then
        \begin{eqnarray}
            U^r_\text{c}
            &=&
            \sqrt{\frac{M}{2r_\text{c}}},
            \\
            v^r_\text{c}
            &=&
            \frac{U^r_\text{c}}{\sqrt{1-3\left[U^r_\text{c}\right]^2}}
            \\
            T_\text{c}
            &=&
            \frac{q\left[v_\text{c}^r\right]^2}{(q+1)(1-q\left[v_\text{c}^r\right]^2)},
        \end{eqnarray}
        as these allow us to find the expressions
        \begin{eqnarray}
            C_1
            &=&
            U_\text{c}^rr_\text{c}^2T_\text{c}^q,
            \\
            C_2
            &=&
            \left[1+(q+1)T_\text{c}\right]^2\left\{1-\frac{2M}{r_\text{c}}+\left[U^r_\text{c}\right]^2\right\}.
        \end{eqnarray}
        It is important to remark that this solution was obtained assuming a Schwarzschild metric. 
        To set up this test we consider a central object with $M=1$ and $a=0$, enabling the hydrodynamics in combination with the Kerr metric module. 
        As initial condition we started from the analytic solution described previously. 
        In order to map the spherically symmetric density distribution we placed the particles uniformly within a unitary sphere following a hexagonal close-packed configuration. 
        Then, we stretched the distribution radially so that they follow the analytic density profile roughly from $r=2.1~\text{M}$ to $r=100~\text{M}$. 
        The simulation utilised $\sim$6 million particles and was run up to a total time of $t=400~\text{M}$. 
        The timestep was set adaptively using the expressions shown in Section~\ref{sec:dt}.
        In order to avoid small timesteps particles that moved inside a sphere of radius $r=1~\text{M}$ were removed from the simulation.

        The results of the simulation at $t=350~\text{M}$ are shown in Figure~\ref{fig:bondi}. 
        Radial profiles of local rest-frame density, radial velocity, smoothing length, and specific internal energy are displayed in top left-, top right-, bottom left- and bottom right-hand side panels, respectively. 
        The solid blue lines represent the analytic solution in steady state, while the orange dots correspond to the values at the particle locations. 
        Note that the agreement between the exact and numerical solutions is excellent for both density and internal energy.
        While the radial velocity agrees overall reasonable well, it also shows noise, which could likely be improved 
        by more careful initial conditions. Note, however, that this velocity noise also appears in the general relativistic version of Phantom \cite[see Figure~12 in][]{liptai2019}, and in the particle-based Finite Volume code GIZMO \citep{hopkins2015}, actually both in the Meshless Finite Volume (MFV) and the Meshless Finite Mass (MFM) variant. 
        Despite this point these results validate the use of the relativistic hydrodynamic and time-independent metric modules coupled.
\section{Stellar tidal disruption in Kerr metric}
\label{sec:application}

    Now we proceed to simulate stellar tidal disruptions in Kerr metric. 
    First, we describe the procedure to compute the initial conditions, i.e. a stellar structure in hydrostatic equilibrium, in Section~\ref{sec:ics}. 
    Then, we present the numerical setup and the analysis methodology in Sections~\ref{sec:models} and~\ref{sec:analysis}, respectively.

    \subsection{Initial conditions}
    \label{sec:ics}
        To obtain good numerical initial conditions for a polytropic star, we follow a three-step procedure: i) we first place particles uniformly in a sphere, ii) perform some radial stretching and iii) further regularise the particle distribution with the ``Artificial Pressure Method" \citep[APM;][]{rosswog2020,rosswog2021}.
        The initial particle sampling consisted of placing the set of particles within a unitary sphere following a hexagonal close-packed configuration. 
        Then, the particles were remapped into the volume of the star so that they follow the desired density profile through the method used in \cite{rosswog2009}. 
        In order to improve the particle distribution so that it is as free as possible of sampling artifacts specific axes (due to the initial hexagonal lattice) and provides a good interpolation accuracy we applied the APM. 
        This approach consists of updating the particle positions due to an artificial pressure $\pi$ that is determined by the relative errors between the current and the desired density distribution $\rho^P$, and it is given by
        \begin{equation}
            \pi_a = \max\left\{1+\frac{\rho_a-\rho^P(\vec{r}_a)}{\rho^P(\vec{r}_a)},0.1\right\}.
        \end{equation}
        Modelled after the hydrodynamic momentum equation, we push the particles in a direction so that their density error decreases until they are located in an optimal position.
        The displacement of a particle $a$ is calculated as
        \begin{equation}
            \delta\vec{r}_a^\text{APM}=-\frac{1}{2}h_a^2 m_0\sum_b\frac{\pi_a+\pi_b}{\rho^*_b}\nabla_aW_{ab}(h_a).
        \end{equation}
        Additionally, an extra term to improve the interpolation accuracy, and the local regularity of the distribution is added, 
        \begin{equation}
            \delta\vec{r}^\text{reg}_a=h_a^4\sum_bW_{ab}(h_a)\vec{\hat{e}}_{ab},
        \end{equation}
        where $\vec{\hat{e}}_{ab}=\left(\vec{r}_a-\vec{r}_b\right)/|\vec{r}_a-\vec{r}_b|$. 
        Then, the complete shift in the particle positions is calculated as
        \begin{equation}
            \delta\vec{r}_a^\text{total}=(1-\zeta)\delta\vec{r}_a^\text{APM}+\zeta\delta\vec{r}_a^\text{reg},
        \end{equation}
        where $\zeta$ is a factor that controls the trade-off between the two conditions that here is set to $\zeta=0.02$. 
        The maximum displacement of the particles per iteration is limited to a fraction $f_h=0.01$ of their smoothing length.
        Thus, the updated position of a particle $a$ is 
        \begin{equation}
            \vec{r}_a\to\vec{r}_a+\min\left\{|\delta \vec{r}_a^\text{total}|,f_hh_a\right\}\vec{\hat{e}}_{\delta r_a},
        \end{equation}
        where we have introduced the magnitude and direction of this displacement  $\delta\vec{r}_a^\text{total}=\delta r_a^\text{total}\vec{\hat{e}}_{\delta r_a}$. 
        For a full derivation of the method we refer the reader to \cite{rosswog2020} and \cite{rosswog2021} for the Newtonian and the general relativistic cases, respectively. 
        
        The resulting particle distribution is later relaxed through evolving the self-gravitating fluid in a Minkowski spacetime. 
        The dissipation terms are switched off but an additional velocity dependent damping force is added in the canonical momentum equation,
        \begin{equation}
            f_\text{damp}^i=-10\frac{v^i}{t_\text{dyn}},
        \end{equation}
        where $t_\text{dyn}$ is the dynamical timescale of the star. 
        This force aids to damp particle motion, so that the star achieves hydrostatic equilibrium. 
        The relaxation simulation is run for at least $20t_\text{dyn}$. 
        \begin{figure}
            \centering
            \includegraphics[width=\linewidth]{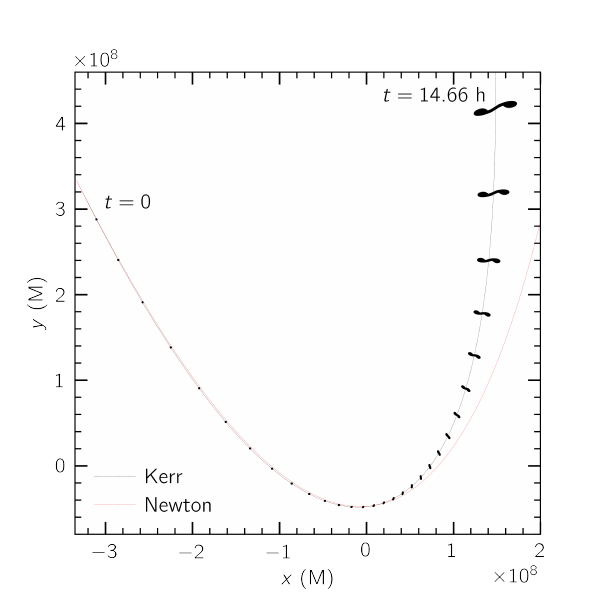}
            \caption{
            Evolution of the stellar disruption for an event with $\beta=1$ and $a=0$ during the first pericentre passage. 
            The location of the SPH particles are shown on the orbital plane spanning from the initial time up to $t=14.66~\text{h}$. 
            Notice the star has a counter-clockwise orbital motion. 
            As a consistency check, the geodesic of a parabolic orbit is shown as a solid grey line. 
            Note the relevance of the relativistic effects by contrasting the geodesic with the Newtonian orbit shown as a solid red line.
            }
            \label{fig:beta1}
        \end{figure}
        \begin{table}
            \centering
            \begin{threeparttable}
            \caption{Simulation runs and parameters. 
            All models considered a Solar-like star with $m_*= 1~\text{M}_{\odot}$ and $R_*=1~\text{R}_{\odot}$ modelled as a polytrope with $n=1.5$ (or {\bf $\Gamma=5/3$}) using 4,194,304 particles.}
            \begin{tabular}{lcccc}
                \hline
                \\
                Simulation
                &
                $M_\bullet$~(M$_{\odot}$)
                &
                $a$/M$_\bullet$
                &
                $\beta$
                &
                $t_\text{final}$~(d)
                \\
                (1) & (2) & (3) & (4) & (5)
                \\
                \hline
                \\
                a0\_$\beta$1 & $10^6$ & 0 & 1 & 3.5
                \\
                a0\_$\beta$2 & $10^6$ & 0 & 2 & 3.5 
                \\
                a0\_$\beta$4 & $10^6$ & 0 & 4 & 3.5 
                \\
                a0\_$\beta$6 & $10^6$ & 0 & 6 & 3.5 
                \\
                a0\_$\beta$8 & $10^6$ & 0 & 8 & 3.5 
                \\
                a0\_$\beta$10 & $10^6$ & 0 & 10 & 0.29$^a$
                \\
                \\
                ap\_$\beta$1 & $10^6$ & +0.99 & 1 & 3.5
                \\
                ap\_$\beta$2 & $10^6$ & +0.99 & 2 & 3.5 
                \\
                ap\_$\beta$4 & $10^6$ & +0.99 & 4 & 3.5 
                \\
                ap\_$\beta$6 & $10^6$ & +0.99 & 6 & 3.5 
                \\
                ap\_$\beta$8 & $10^6$ & +0.99 & 8 & 3.5 
                \\
                ap\_$\beta$10 & $10^6$ & +0.99 & 10 & 3.5
                \\
                \\
                am\_$\beta$1 & $10^6$ & -0.99 & 1 & 3.5
                \\
                am\_$\beta$2 & $10^6$ & -0.99 & 2 & 3.5 
                \\
                am\_$\beta$4 & $10^6$ & -0.99 & 4 & 3.5 
                \\
                am\_$\beta$6 & $10^6$ & -0.99 & 6 & 3.5 
                \\
                am\_$\beta$8 & $10^6$ & -0.99 & 8 & 0.29$^a$ 
                \\
                am\_$\beta$10 & $10^6$ & -0.99 & 10 & 0.29$^a$
                \\
                \hline
                \hline
            \end{tabular}
            \label{tab:sims}
            \begin{tablenotes}
                \item
                \textit{Notes.} 
                Column~1: simulation name. 
                Column~2: mass of the central black hole. 
                Column~3: spin of the central black hole.
                Column~4: impact strength $\beta=r_\text{t}/r_\text{p}$. 
                Column~5: total simulation time.
                \\
                $^a$Simulations were stopped earlier due to prohibitive timestep size.
            \end{tablenotes}
            \end{threeparttable}
        \end{table}
    \subsection{Numerical models}
    \label{sec:models}
        In this work, we focus on the disruption of a Solar-like star, i.e. with mass 1~M$_{\odot}$ and radius 1~R$_{\odot}$ by a black hole of mass $M_\bullet=10^6~\text{M}_{\odot}$ and spin $a/M_\bullet=-0.99,~0,~+0.99$. 
        The star is modelled as a polytrope with index $n=1.5$ sampled with 4,194,304 SPH particles. 
        The star is placed on an equatorial parabolic orbit at an initial distance of $10~r_\text{t}$ from the black hole. 
        The pericentre distance is an input parameter specified by the  penetration factor $\beta$, which was chosen as $\beta=1,~2,~4,~6,~8,~10$.
        The star is initialised in the second quadrant with the black hole at the origin, so that the star travels in a counter-clockwise (positive) manner. 

        The initial position and velocity for placing the star on a parabolic motion were computed by calculating the $z$ component of the specific angular momentum $\ell_z$ (see Appendix~\ref{sec:parabolic}) and by using the orbital specific energy $E=1$. 
        With these conserved quantities, it is straightforward to calculate the Carter constant $\mathcal{Q}$, and the specific angular momentum $\ell$. 
        Then, we used the four conserved quantities to compute the initial position and velocity in CKS coordinates following \cite{tejeda2017}.
            
        Figure~\ref{fig:beta1} shows as an example the evolution of the position $(x,y)$ of the SPH particles for the case of a Schwarzschild SMBH ($a=0$) and a ``grazing impact" ($\beta=1$). 
        Notice how at $t=0$ the star is a round (dot-like) structure that becomes elongated due to the tidal forces of the central object, especially after the closest encounter. 
        As expected, the motion of the structure follows the geodesic (see dashed grey line) describing a parabolic motion. 
        Note that, especially after pericentre passage, the trajectory deviates significantly from the Newtonian parabolic orbit (see solid red line).
        All simulations were run up to $t=3.5$~d, and at this point they were analysed. 
        In total, we ran 18 simulations whose computational cost was about 60,000-80,000 CPU hours for each depending of the exact parameters used.
        A summary with the parameters of each run is provided in Table~\ref{tab:sims}. 
    \subsection{Analysis}
    \label{sec:analysis}
\begin{figure}
            \centering
            \includegraphics[width=0.925\linewidth]{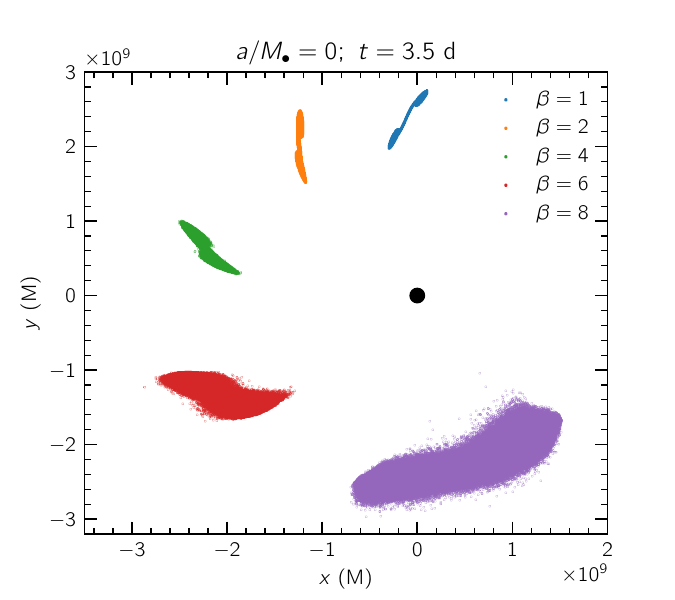}
            \includegraphics[width=0.925\linewidth]{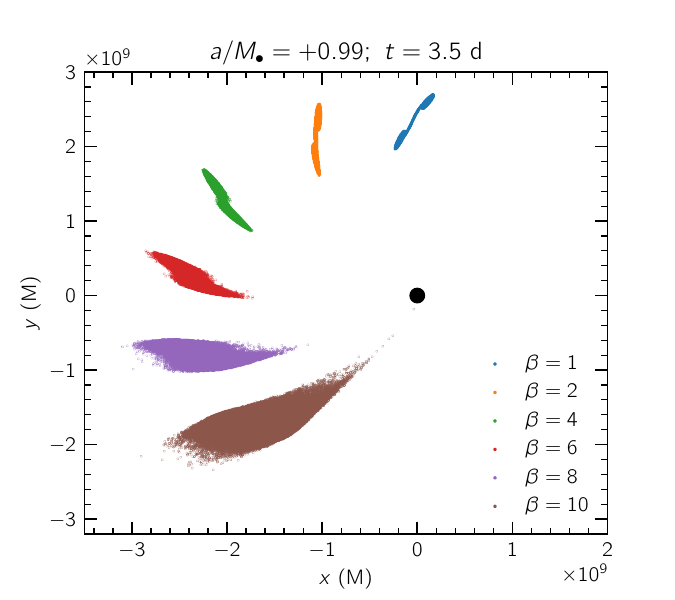}
            \includegraphics[width=0.925\linewidth]{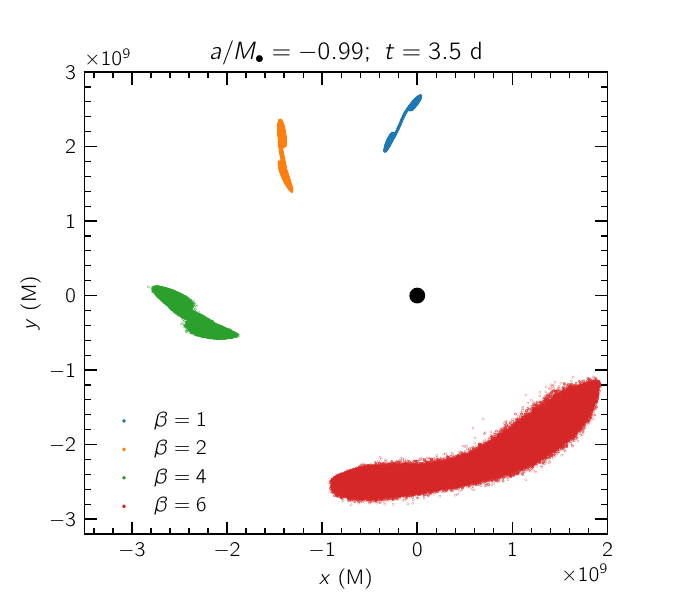}
            \caption{
            Projections of the SPH particles on the $xy$ plane at the final time $t=3.5$~d.
            The panels show the models with $a/M_\bullet=0,~+0.99,~-0.99$ in the top, middle, and bottom, respectively. 
            The cases $\beta=1,2,4,6,8,10$ are represented in blue, orange, green, red, purple, and brown colours, respectively. 
            The central black hole location is represented as a black dot (size not to scale). 
            Notice that all panels show the same extension of the $xy$ plane.
            }
            \label{fig:final_state}
        \end{figure}

        The final state of every simulation run is shown in Figure~\ref{fig:final_state}. 
        Here, we show the distribution of the SPH particles projected on the $xy$ plane at the final simulation time. 
        The top, middle, and bottom panels show the models with $a/M_\bullet=0,+0.99,-0.99$, respectively. 
        In addition, the impact strength $\beta=1,2,4,6,8,10$ are represented in blue, orange, green, red, purple, and brown colours, respectively. 
        At this point, we analysed the state of the disrupted star. 
        First, we estimated the mass fallback rate directly from the simulation output following the same method as in   \cite{gafton2019}. 
        We used the properties of each particle to compute the conserved variables of motion $E,\ell_z,\mathcal{Q},\ell^2$. 
        Then, we calculated the radial turning points of their geodesics finding the roots of the polynomial expression of the radial equation in Kerr spacetime \citep[see equation A18 in Appendix A2;][]{tejeda2017}. 
        We identified the two largest roots as the apo- and pericentre of the orbits, while discarding the unbound particles. 
        Then, we proceeded to integrate the time coordinate along their geodesic from the current state until the next pericentre, that is, taking into account the time coordinate to pericentre for infalling orbits while considering the total of the motion to apocentre and then to pericentre for outwards orbits. 
        This procedure assigns a time coordinate to each particle at the next pericentre passage. 
        A time histogram is constructed from this calculation weighing each particle for its mass divided by bthe time bin width. 
        The outcome is the function $\dot{M}_\text{fb}(t)$ that corresponds to the mass fallback rate. 

        In order to make comparisons with previous studies \citep[e.g.][]{gafton2019,jankovic2023}, we have extracted the characteristic properties of the fallback rate function: its maximum value $\max\{\dot{M}_\text{fb}(t)\}$, time of the maximum $t_\text{peak}$, late-time decay exponent $n_\infty$, and the characteristic rise-to-peak timescale $\tau_\text{rise}$. 
        The exponent $n_\infty$ was calculated via fitting a decay $\propto$~$t^{n_\infty}$ to the portion of the function that satisfies $\dot{M}_\text{fb}(t)>0.1\times\max\{\dot{M}_\text{fb}(t)\}$, where $t>t_\text{peak}$. 
        The rise to peak timescale was obtained fitting a Gaussian function $\propto \exp\{-(t-t_\text{peak})^2/[2\tau_\text{rise}^2]\}$ when $0.2\times\max\{\dot{M}_\text{fb}(t)\}<\dot{M}_\text{fb}(t)<0.8\times\max\{\dot{M}_\text{fb}(t)\}$, where $t<t_\text{peak}$. 
\section{Results}
\label{sec:results}
    We proceed to describe the final outcome of the numerical simulations, and we explore in particular the impact of the penetration factor $\beta$ and the black hole spin $a$. 
    We further calculate mass
    fallback rates, and their characteristic properties to compare them across our explored models. 
    \subsection{The impact of penetration for non-spinning black hole}
        All simulations were run up to $t=3.5$~d which corresponds to a point after the first pericentre passage but before the next one. 
        It is relevant to remark that all models resulted in fully disrupted stars, as expected for polytropes with $n=1.5$ and $\beta\gtrsim1$ which also has been confirmed by previous works \citep[e.g.][]{guillochon2013,gafton2019}.
        As shown in the top panel of Figure~\ref{fig:final_state}, the stellar streams at the end of the simulation have different structures depending on the impact strength. 
        First, it can be seen that with increasing impact strength the structure that connects both lobes of the disrupted star becomes less relevant. 
        Specifically, at $\beta\gtrsim4$ this feature disappears as it is related to the relevance of the self-gravity. 
        The deeper the encounter, the self-gravity becomes less important, so the stream evolves effectively solely under the effect of the curved metric. 
        Unfortunately, the case with $\beta=10$ was not possible to simulate further than $t=0.29$~d, which corresponds to the first pericentre passage, due to particles approaching too close to the central object.
        In analogous previous works, such a behaviour has also been observed due to particles moving on plunging orbits \citep[e.g.][]{jankovic2023}. 
        In our case, the Kerr metric in CKS coordinates should not have this problem as the only singularity is located at the origin. 
        Yet, our simulations also presents numerical problems as the timestep decreases continuously causing the simulation to effectively stall due to particles being too close to the singularity.
    \subsection{The role of extreme black hole spin}
        Observing the middle and bottom panels of Figure~\ref{fig:final_state}, it is possible to see changes in the location and structures due to the impact of the prograde and retrograde spinning black holes, respectively. 
            
        First, the exact location of the stream at the end of the simulation is affected by precession, especially in deep events and if the black hole is spinning. 
        This is why even for the same impact parameter (represented with the same colour) the final location of the structure is different across the panels of Figure~\ref{fig:final_state}.
            
        Second, since the black rotation affects the net effect of the tidal field on the star, the prograde cases deal with a slightly weaker tidal field. 
        This can be seen especially on the fact that all simulations up to $\beta=10$ managed to reach their final time. 
        On the contrary, in the retrograde cases the stars are exposed to a stronger tidal field which translated into capturing the complete simulations only up to $\beta=6$. 
        In these numerically problematic cases, again it was not possible to run the models further than $t=0.29~\text{d}$. 
        Overall, the final stellar stream structures are relatively similar for $\beta\lesssim4$, even across the non-spinning SMBH case. 
        In deeper cases, the effects of the spin makes a clear difference affecting the stream dramatically in the retrograde case giving it a fully blob-like shape. 
        Meanwhile in the prograde cases, since the tidal force is milder it still can be seen the stretch exerted by the black hole on the streams.
        These features are a direct consequence of the particle energy spread after the pericentre passage is narrower for prograde motion and wider in the case of retrograde motion. 
        \begin{figure}
            \centering
            \includegraphics[width=0.925\linewidth]{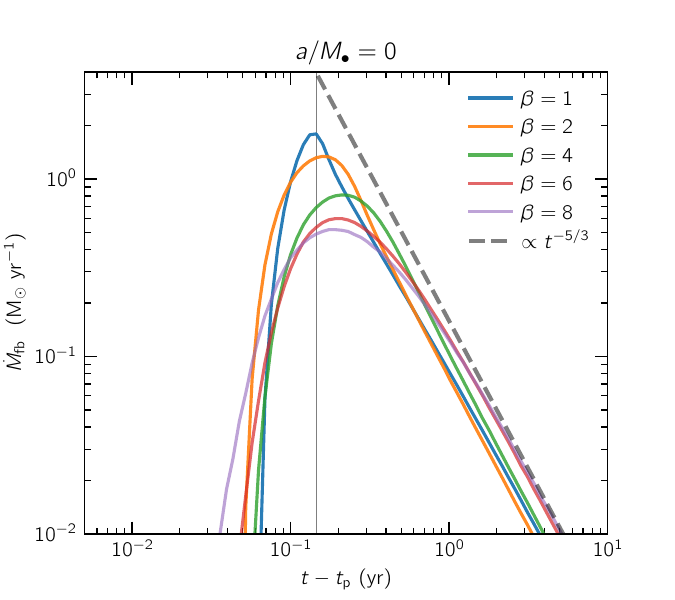}
            \includegraphics[width=0.925\linewidth]{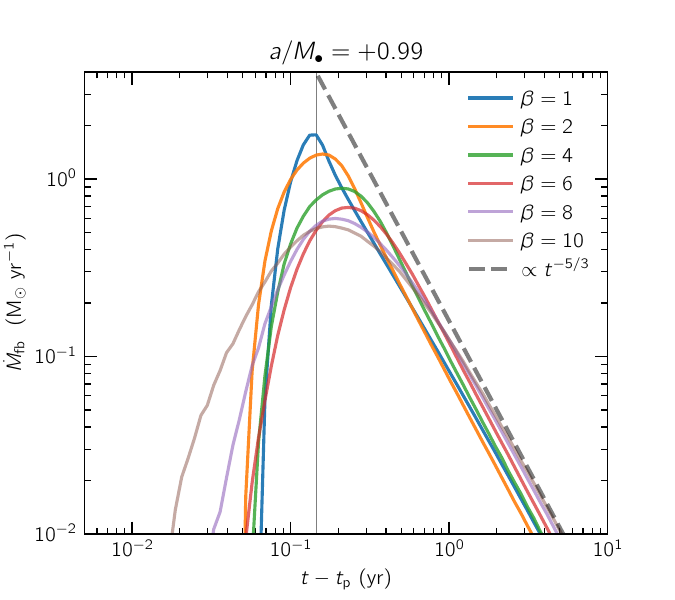}
            \includegraphics[width=0.925\linewidth]{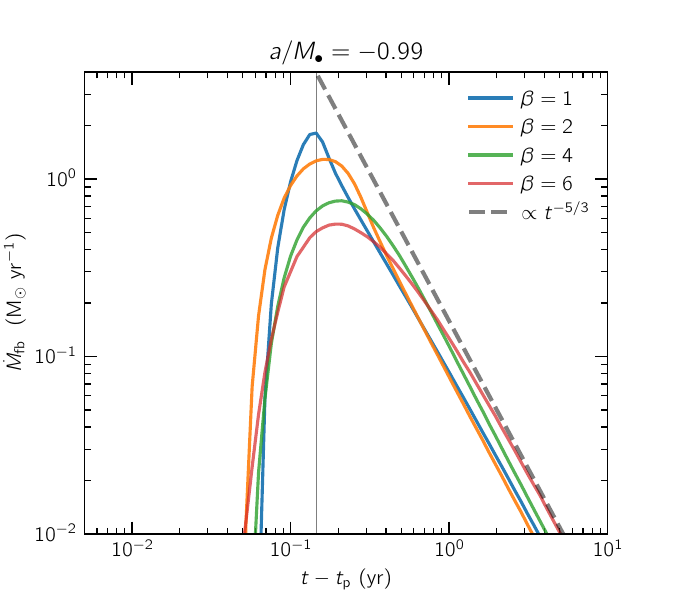}
            \caption{
            Mass fallback rates as a function of time from pericentre passage for events with impact strength $\beta=1,~2,~4,~6,~8,~10$. 
            The top, middle, and bottom panels show the cases for black holes with with $a/M_\bullet=0,+0.99,-0.99$, respectively. 
            The vertical solid grey line highlights the location of the peak of the case $\beta=1$ for guiding the eye.
            The dashed grey line represents a decay $\propto t^{-5/3}$ shown as reference.
            }
            \label{fig:fallbacks}
        \end{figure}
        \begin{figure*}
            \centering
            \includegraphics[width=0.45\linewidth]{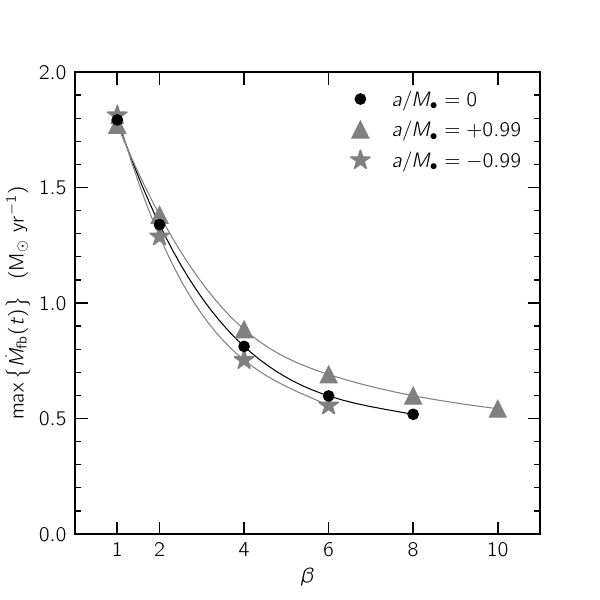}
            \includegraphics[width=0.45\linewidth]{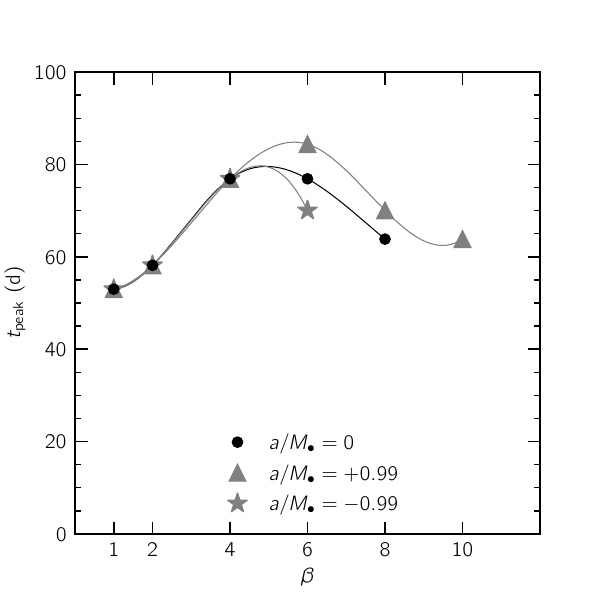}
            \includegraphics[width=0.45\linewidth]{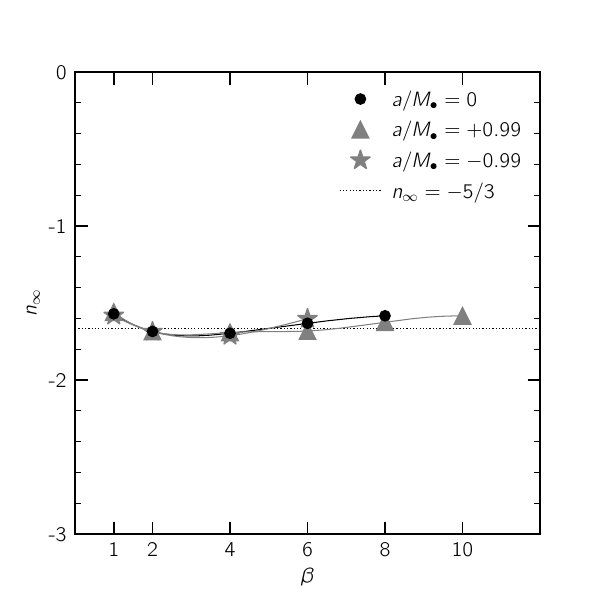}
            \includegraphics[width=0.45\linewidth]{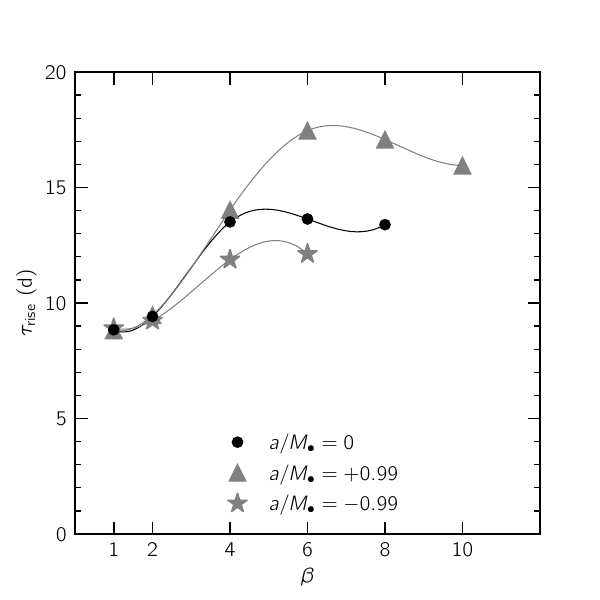}
            \caption{
            Maximum mass fallback rate, time of the maximum, late-time power-law decay exponent, and rise to peak timescale as functions of the penetration factor are shown on the top left-, top right-, bottom left-, and bottom right-hand side panels, respectively. 
            Every single data point represent a different simulation run. 
            Black dots, grey triangles, and grey stars represent the cases with spins $a/M_\bullet=0,+0.99,-0.99$, respectively. 
            The solid lines connecting the data points correspond to B-spline fits whose knots and coefficients are presented in Table~\ref{tab:fits}.
            }
            \label{fig:max_fallback}
        \end{figure*}
    \subsection{Fallback rates}
    \label{sec:fallback}
        Figure~\ref{fig:fallbacks} shows the fallback rates as functions of time from the pericentre passage. 
        The top, middle, and bottom panels correspond to the results of the non-spinning, the rapidly prograde ($a/M_\bullet=+0.99$), and retrograde spinning ($a/M_\bullet=-0.99$) black hole cases, respectively. 
        Since some cases ($\beta\ge8$) with a rapidly spinning retrograde case lead to the plunging of a significant fraction of the stellar stream these were not included in this analysis. 
        In general, the fallback mass rates do not differ dramatically. 
        However, it can be seen that in all cases the amplitude of the peak decreases with impact strength as well as all rates tend to follow the analytical $\propto t^{-5/3}$ at long timescales ($\gtrsim 1$~yr). 
        Given that this power law is a result of Kepler's third law, it is not surprising that this result is recovered. 
            
        We performed a quantitative analysis of the following properties to characterise the mass fallback rates: $\max\{\dot{M_\text{fb}}\}$, $t_\text{peak}$, $n_\infty$, and $\tau_\text{rise}$. 
        Figure~\ref{fig:max_fallback} shows each of these properties as a function of the impact strength $\beta$. 
        Each data point in these panels corresponds to a single simulation run. 
        Additionally, we fitted B-spline functions to the sets of simulations with a given spin value inspired by previous work \citep[e.g.][]{guillochon2013,gafton2019}, so that the outputs can be used for other scopes. 
        The exact fitting knots and coefficients can be found in Table~\ref{tab:fits} in Appendix~\ref{app:fitting}. 
        Starting from the top left-hand side panel, it can be seen that indeed the mass fallback rate peak decreases with increasing impact strength. 
        This is a result of the tidal field becoming stronger with impact strength which causes a wider energy spread and a more stretched, diluted stream returning to pericentre. 
        The black hole spin makes this effect stronger, as retrograde orbits \citep[with a stronger net tidal field; see, e.g][]{cocco2026} cause an even wider energy spread. 
        This explains the peaks becoming slightly smaller with impact strength. 
        The contrary applies for prograde orbits where the peak values are slightly larger.
        Second, in the top right-hand side panel, the time to reach the peak as a function of impact strength  is shown. 
        At small impact strength ($\beta\lesssim4$), this timescale increases proportionally with impact strength  as a result of the stronger tidal field creating a wider energy dispersion after the first pericentre passage. 
        Only at higher impact strength the trend reverses and the time to reach the peak becomes shorter with deeper encounters. 
        Our analysis shows that this might be a consequence of the self-gravity being completely negligible in deep encounters. 
        Then, the time of the peak is completely determined by the metric and the strength of the tidal field. 
        This is also corroborated by  the effect of the black hole spin. 
        Note that only in deep encounters the spin affects the time of the peak, as again the effects of the frame dragging increases (decreases) the tidal field in the case of prograde (retrograde) motion compared to the non-spinning case.
            
        In the bottom left-hand side panel, as we had seen by eye, all runs irrespective of impact strength or spin show that the mass fallback rate decays with $t^{-5/3}$ over long timescales. 
        This fact is expected and serves as consistency check with previous work and analytic predictions \citep[e.g.][]{rees1988,ramirezruiz09,guillochon2013}.
        Last, in the bottom  right-hand side panel, $\tau_\text{rise}$ is shown as a function of impact strength. 
        Initially, the rise-to-peak timescale increases with impact strength but the trend tends to either saturate or reverse around $\beta\sim4-6$, in the same manner as in the time of the peak. 
        Thus, we attribute this behaviour to the same mechanism. 
        In shallow events ($\beta\lesssim4$), self-gravity plays a small role in keeping the energy dispersion of the particles slightly narrower, so the stream is denser and the rise steeper. 
        Only in the case of deep events ($\beta\gtrsim4$), the rise-to-peak timescale tends to saturate as it is only determined by the metric and the self-gravity effects are completely negligible. 
        In this regime, we also observe that the black hole spin can delay (shorten) the rise if the star moves in a prograde (retrograde) manner.
\section{Discussion}
\label{sec:disc}
    \subsection{Comparison with previous work}
        In the literature, there has been work done simulating globally the hydrodynamics of stellar tidal disruptions considering general relativistic effects \citep[e.g.][]{cheng2014,tejeda2017,gafton2019,liptai2019b,jankovic2023,ryu2023,price2024,chan2026}. 
        However, only few of them have systematically examined the role of both the impact strength and black hole spin.
        Thus, we concentrate our comparisons on \cite{gafton2019} and \cite{jankovic2023}.
        The former used the code derived from MAGMA \citep{rosswog2007}, where the Newtonian accelerations are replaced by the ``Generalised Newtonian" description developed in \cite{tejeda2017} which allows to include general relativistic effects.  
        Since they also considered polytropic stars with $n=1.5$ the comparison can be done directly.
        The later used the code Phantom \citep{price2018} which employs, as our approach, SPH in Kerr spacetime, but they considered stellar structures computed with the MESA code \citep{paxton2011} rather than polytropic models. 
        In fact, \cite{jankovic2023} investigated a variety of stellar structures, so in order to make a sensible comparison with their work, we focus on their model of a zero age main-sequence star of 1~M$_{\odot}$.
        It is relevant to add that the numerical resolution to sample the stars in these works were lower, being $\sim$200k in \cite{gafton2019} and $\sim$1M in \cite{jankovic2023}; while here we have employed $\sim$4M particles. 

        In general, we do recover the same trends found by \cite{gafton2019} and \cite{jankovic2023}, namely $\max\{\dot{M}_\text{fb}(t)\}$, $n_\infty$, $t_\text{peak}$, and $\tau_\text{rise}$ as functions of $\beta$. 
        More specifically, our models agree with seeing a decline in the mass fallback rate peaks with increasing impact strength. 
        Even more, we see the exact slight increase (decrease) on these peaks if the star moves on a prograde (retrograde) orbit. 
        For the timescales as functions of impact strength, we do observe similarities, but also some differences on part of the trends by \cite{jankovic2023}. 
        We attribute these differences to  using more realistic stellar structures rather than the polytropic ones of this work. 
        Last, the agreement  can also be seen in the long-term power-law decay. 
        However, we do observe noticeable quantitative differences and actually different trends in deep events compared to \cite{gafton2019}. 
        For instance, the fallback rate of the fiducial case, $\beta=1$ and $a=0$, peaks around $\sim$1.3~M$_{\odot}$~yr$^{-1}$, but the case $\beta=4$ has the minimum peak at $\sim$0.6~M$_{\odot}$~yr$^{-1}$ which  increases with impact strength. 
        Our results show that the fiducial case peaks at $\sim$1.8~M$_{\odot}$~yr$^{-1}$ and declines monotonically with $\beta$ reaching $\sim$0.5~M$_{\odot}$~yr$^{-1}$ for $\beta=8$. 
        \cite{jankovic2023} also see this monotonic decline with impact strength which they attribute to differences in code implementations, and not to the use of different stellar structures. 
        However, since our approach is similar to \cite{jankovic2023} and we do observe the same differences, it is likely that code differences play a role, but also the significant resolution difference may contribute. 
        Differences compared to \cite{gafton2019} are also seen in the time of the peak: the fiducial case shows $t_\text{peak}=60$~d, reaches its maximum at $\sim$65~d for $\beta=$2, and then decays to $\sim$30~d for $\beta=10$. 
        Meanwhile our results points to $t_\text{peak}=55$~d in the fiducial case, $\sim$80~d at $\beta=4$, and $\sim$65~d at $\beta=8$. 
        Unfortunately, it is not possible to do sensible quantitative comparisons with \cite{jankovic2023} as the stellar structure affects the exact values of these quantities. 
        In forthcoming work, we will explore more realistic stellar structures to assess quantitatively the potential difference with other state-of-the-art models.
    \subsection{Longer-term modelling and challenges}
        We have simulated tidal stellar disruptions on parabolic orbits for a total of 3.5~d. 
        Observationally, TDEs have been followed for months to years timescales. 
        Thus, ideally we would like to simulate the stellar tidal interaction over comparable timescales. 
        The big computational challenge lies in capturing the second pericentre passage with high-enough resolution in order to resolve the nozzle shock properly. 
        To date, there are only a couple of works that have managed to do this successfully with a single global simulation. 
        \cite{hu2026} made use of a module that dynamically splits SPH particles to enhance the resolution locally based on geometric criteria in order to resolve the nozzle shock and avoid the numerical \textit{fanning} of the stream \citep[e.g.][]{price2024} due to finite resolution effects.
        Another solution is using of GPU-accelerated SPH simulations to use as many particles as $\sim$10$^{10}$ to ensure resolving the stream during the second pericentre \citep{kubli2026}. 
        These studies also underline that effects due to finite numerical resolution are still significant in most of today's models. 
        Attempting to push our models to such timescales demands the development of at least one of these alternatives which we will implement in our code in the future.
\section{Conclusions}
\label{sec:conclusions}
    Throughout this work, we have introduced a new tool to perform hydrodynamic modelling of stellar tidal disruptions in curved spacetime using the code SPHINCS. 
    We have implemented and validated a  Kerr metric module, together with Newtonian self-gravity of the stellar fluid. 
    For the applications we have in  mind, TDEs, this is an excellent approach, since the fluid's self-gravity only causes a tiny perturbation in the metric generated by the much more massive black hole. 
    In addition, we have added the option of evolving the (pseudo-)entropy instead of the canonical energy to follow the dynamics of the fluid in a curved spacetime. 
    With this new tool, we have conducted simulations of stellar tidal disruptions of Solar-like stars by both non-spinning and rapidly rotating SMBHs. 
    We performed 18 simulations considering a star sampled by $\sim$ 4M SPH particles at a reasonable computational cost: $\sim$900k CPU hours in total. 
    This fact has permitted the exploration of a wide range of parameters, namely the impact strength $\beta=1,2,4,6,8,10$ and the spin of the black hole $a/M_\bullet=-0.99,0,+0.99$. 
    We have constructed mass fallback rates from each of the simulations, and derived some characteristic quantities from them. 
    We contrasted the trends and properties of these rates with state-of-the-art simulations using other  approaches and we found reasonable agreement. 
    Mass fallback rates from stellar disruptions by Kerr black holes result in lower peaks, they reach their peak height faster, and they are wider with increasing impact strength, and vice-versa. 
    If the black hole is rapidly spinning and the star moves on a prograde (retrograde) orbit, the peak, time of the peak, and rise-to-peak timescale increase (decrease) further.
    As expected, the fallback rates converge to the characteristic $t^{-5/3}$ on long timescales ($\gtrsim$1~yr), provided that the star is fully disrupted.

    Having set a benchmark for stellar tidal disruptions by a (non-)spinning SMBH, we plan to make use of this numerical tool to perform a wide variety of simulations at different scales. 
    Further efforts will be invested in more realistic stellar models, different SMBH masses as well as off-equatorial orbits. 
    A non-zero inclination between the orbital plane and the black hole spin direction can create precession off the orbital plane. This may lead to 
    the tidal stream wrapping around the black hole many times before finally colliding with itself, which may trigger an observable TDE event \citep{guillochon2015}.
    The impact of this phenomenon on TDE light curves has been quantified \citep{calderon2024,calderon2026}, but the simulations were two dimensional azimuthally symmetric and the initial conditions were computed semi-analytically. 
    Now, using SPHINCS it is possible to create more realistic initial conditions to quantify appropriately whether or not the spin of the SMBH can be imprinted in TDE light curves. 
    Other applications are along the lines of partial stellar disruptions of a brown dwarf, inspired on the extremely high mass ratio inspirals that Laser Interferometer Space Antenna (LISA) might capture in the future \citep[e.g.][]{amaro2025}.  
\section*{Data availability}
    The output files from our simulations will be shared on reasonable request to the corresponding author.
\begin{acknowledgements}
    We would like to thank Dr. Taj Jankovi\v{c} for very helpful discussions and Dr. Raghav Arora for double checking the analytic calculation of the derivatives of the Kerr metric.
    The research of DC and SR has been supported by the Deutsche Forschungsgemeinschaft (DFG, German Research Foundation) under Germany’s Excellence Strategy - EXC 2121 - ``Quantum Universe” - 390833306. 
    The research of DC has been funded by the Alexander von Humboldt Foundation. 
    DC also acknowledges the financial support from ANID-FONDECYT Regular 1251444.
    SR has been supported by the Swedish Research Council (VR) under grant number 2020-05044, by the research environment grant “Gravitational Radiation and Electromagnetic Astrophysical Transients” (GREAT) funded by the Swedish Research Council (VR) under Dnr 2016-06012, by the Knut and Alice Wallenberg Foundation under grant Dnr. KAW 2019.0112, and by the European Research Council (ERC) Advanced Grant INSPIRATION under the European Union’s Horizon 2020 research and innovation programme (Grant agreement No. 101053985). 
    The numerical simulations of this work were run on the high-performance computing system Viper of the Max Planck Computing and Data Facility. 
    This work made use of \textsc{splash} to analyse the SPH simulations \citep{price2007}, as well as of the 
    \textsc{python} libraries \textsc{numpy} \citep{harris2020}, \textsc{matplotlib} \citep{hunter2007}, \textsc{scipy} \citep{virtanen2020}, and the NASA’s Astrophysics Data System.
\end{acknowledgements}

\bibliographystyle{aa} % style aa.bst
\bibliography{tde} % your references Yourfile.bib

@ARTICLE{auchettl17,
       author = {{Auchettl}, Katie and {Guillochon}, James and {Ramirez-Ruiz}, Enrico},
        title = "{New Physical Insights about Tidal Disruption Events from a Comprehensive Observational Inventory at X-Ray Wavelengths}",
      journal = {\apj},
         year = 2017,
        month = apr,
       volume = {838},
       number = {2},
          eid = {149},
        pages = {149},
          doi = {10.3847/1538-4357/aa633b},
archivePrefix = {arXiv},
       eprint = {1611.02291},
 primaryClass = {astro-ph.HE},
       adsurl = {https://ui.adsabs.harvard.edu/abs/2017ApJ...838..149A}
}

@ARTICLE{maguire20,
       author = {{Maguire}, Kate and {Eracleous}, Michael and {Jonker}, Peter G. and {MacLeod}, Morgan and {Rosswog}, Stephan},
        title = "{Tidal Disruptions of White Dwarfs: Theoretical Models and Observational Prospects}",
      journal = {\ssr},
         year = 2020,
        month = mar,
       volume = {216},
       number = {3},
          eid = {39},
        pages = {39},
          doi = {10.1007/s11214-020-00661-2},
archivePrefix = {arXiv},
       eprint = {2004.00146},
 primaryClass = {astro-ph.HE},
       adsurl = {https://ui.adsabs.harvard.edu/abs/2020SSRv..216...39M}
}

@ARTICLE{wevers23,
       author = {{Wevers}, T. and {Coughlin}, E.~R. and {Pasham}, D.~R. and {Guolo}, M. and {Sun}, Y. and {Wen}, S. and {Jonker}, P.~G. and {Zabludoff}, A. and {Malyali}, A. and {Arcodia}, R. and {Liu}, Z. and {Merloni}, A. and {Rau}, A. and {Grotova}, I. and {Short}, P. and {Cao}, Z.},
        title = "{Live to Die Another Day: The Rebrightening of AT 2018fyk as a Repeating Partial Tidal Disruption Event}",
      journal = {\apjl},
         year = 2023,
        month = jan,
       volume = {942},
       number = {2},
          eid = {L33},
        pages = {L33},
          doi = {10.3847/2041-8213/ac9f36},
archivePrefix = {arXiv},
       eprint = {2209.07538},
 primaryClass = {astro-ph.HE},
       adsurl = {https://ui.adsabs.harvard.edu/abs/2023ApJ...942L..33W}
}

@ARTICLE{guillochon2015,
       author = {{Guillochon}, James and {Ramirez-Ruiz}, Enrico},
        title = "{A Dark Year for Tidal Disruption Events}",
      journal = {\apj},
         year = 2015,
        month = aug,
       volume = {809},
       number = {2},
          eid = {166},
        pages = {166},
          doi = {10.1088/0004-637X/809/2/166},
archivePrefix = {arXiv},
       eprint = {1501.05306},
 primaryClass = {astro-ph.HE},
       adsurl = {https://ui.adsabs.harvard.edu/abs/2015ApJ...809..166G}
}

@BOOK{baumgarte2010,
       author = {{Baumgarte}, Thomas W. and {Shapiro}, Stuart L.},
        title = "{Numerical Relativity: Solving Einstein's Equations on the Computer}",
         year = 2010,
         publisher = "Cambridge University Press",
       adsurl = {https://ui.adsabs.harvard.edu/abs/2010nure.book.....B}
}

@ARTICLE{rosswog2026,
       author = {{Rosswog}, Stephan},
        title = "{SPH methods in the modelling of compact objects}",
      journal = {arXiv e-prints},
         year = 2026,
        month = jul,
          eid = {arXiv:2607.14828},
        pages = {arXiv:2607.14828},
          doi = {10.48550/arXiv.2607.14828},
archivePrefix = {arXiv},
       eprint = {2607.14828},
 primaryClass = {astro-ph.HE},
       adsurl = {https://ui.adsabs.harvard.edu/abs/2026arXiv260714828R}
}

@ARTICLE{tanikawa2017,
       author = {{Tanikawa}, Ataru and {Sato}, Yushi and {Nomoto}, Ken'ichi and {Maeda}, Keiichi and {Nakasato}, Naohito and {Hachisu}, Izumi},
        title = "{Does Explosive Nuclear Burning Occur in Tidal Disruption Events of White Dwarfs by Intermediate-mass Black Holes?}",
      journal = {\apj},
         year = 2017,
        month = apr,
       volume = {839},
       number = {2},
          eid = {81},
        pages = {81},
          doi = {10.3847/1538-4357/aa697d},
archivePrefix = {arXiv},
       eprint = {1703.08278},
 primaryClass = {astro-ph.HE},
       adsurl = {https://ui.adsabs.harvard.edu/abs/2017ApJ...839...81T}
}

@ARTICLE{kerr1963,
       author = {{Kerr}, Roy P.},
        title = "{Gravitational Field of a Spinning Mass as an Example of Algebraically Special Metrics}",
      journal = {\prl},
         year = 1963,
        month = sep,
       volume = {11},
       number = {5},
        pages = {237-238},
          doi = {10.1103/PhysRevLett.11.237},
       adsurl = {https://ui.adsabs.harvard.edu/abs/1963PhRvL..11..237K}
}

@ARTICLE{rosswog2009,
       author = {{Rosswog}, S. and {Ramirez-Ruiz}, E. and {Hix}, W.~R.},
        title = "{Tidal Disruption and Ignition of White Dwarfs by Moderately Massive Black Holes}",
      journal = {\apj},
         year = 2009,
        month = apr,
       volume = {695},
       number = {1},
        pages = {404-419},
          doi = {10.1088/0004-637X/695/1/404},
archivePrefix = {arXiv},
       eprint = {0808.2143},
 primaryClass = {astro-ph},
       adsurl = {https://ui.adsabs.harvard.edu/abs/2009ApJ...695..404R}
}

@Article{harris2020, 
    title         = {Array programming with {NumPy}}, 
    author        = {Charles R. Harris and K. Jarrod Millman and St{\'{e}}fan J. van der Walt and Ralf Gommers and Pauli Virtanen and David Cournapeau and Eric Wieser and Julian Taylor and Sebastian Berg and Nathaniel J. Smith and Robert Kern and Matti Picus and Stephan Hoyer and Marten H. van Kerkwijk and Matthew Brett and Allan Haldane and Jaime Fern{\'{a}}ndez del R{\'{i}}o and Mark Wiebe and Pearu Peterson and Pierre G{\'{e}}rard-Marchant and Kevin Sheppard and Tyler Reddy and Warren Weckesser and Hameer Abbasi and Christoph Gohlke and Travis E. Oliphant},
    year          = {2020},
    month         = sep,
    journal       = {Nature},
    volume        = {585},
    number        = {7825},
    pages         = {357--362},
    doi           = {10.1038/s41586-020-2649-2},
    publisher     = {Springer Science and Business Media {LLC}},
    url           = {https://doi.org/10.1038/s41586-020-2649-2}
}

@Article{hunter2007, 
    Author    = {Hunter, J. D.},
    Title     = {Matplotlib: A 2D graphics environment}, 
    Journal   = {Computing in Science \& Engineering},
    Volume    = {9}, 
    Number    = {3}, 
    Pages     = {90--95},
    publisher = {IEEE COMPUTER SOC}, 
    doi       = {10.1109/MCSE.2007.55}, 
    year      = 2007
}

@ARTICLE{price2007,
       author = {{Price}, Daniel J.},
        title = "{splash: An Interactive Visualisation Tool for Smoothed Particle Hydrodynamics Simulations}",
      journal = {\pasa},
         year = 2007,
        month = oct,
       volume = {24},
       number = {3},
        pages = {159-173},
          doi = {10.1071/AS07022},
archivePrefix = {arXiv},
       eprint = {0709.0832},
 primaryClass = {astro-ph},
       adsurl = {https://ui.adsabs.harvard.edu/abs/2007PASA...24..159P}
}

@ARTICLE{liptai2019,
       author = {{Liptai}, David and {Price}, Daniel J.},
        title = "{General relativistic smoothed particle hydrodynamics}",
      journal = {\mnras},
         year = 2019,
        month = may,
       volume = {485},
       number = {1},
        pages = {819-842},
          doi = {10.1093/mnras/stz111},
archivePrefix = {arXiv},
       eprint = {1901.08064},
 primaryClass = {astro-ph.IM},
       adsurl = {https://ui.adsabs.harvard.edu/abs/2019MNRAS.485..819L}
}

@ARTICLE{lupi2023,
       author = {{Lupi}, Alessandro},
        title = "{A general relativistic extension to mesh-free methods for hydrodynamics}",
      journal = {\mnras},
         year = 2023,
        month = feb,
       volume = {519},
       number = {1},
        pages = {1115-1131},
          doi = {10.1093/mnras/stac3574},
archivePrefix = {arXiv},
       eprint = {2210.05682},
 primaryClass = {gr-qc},
       adsurl = {https://ui.adsabs.harvard.edu/abs/2023MNRAS.519.1115L}
}

@ARTICLE{rosswog2021,
       author = {{Rosswog}, S. and {Diener}, P.},
        title = "{SPHINCS\_BSSN: a general relativistic smooth particle hydrodynamics code for dynamical spacetimes}",
      journal = {Classical and Quantum Gravity},
         year = 2021,
        month = jun,
       volume = {38},
       number = {11},
          eid = {115002},
        pages = {115002},
          doi = {10.1088/1361-6382/abee65},
archivePrefix = {arXiv},
       eprint = {2012.13954},
 primaryClass = {gr-qc},
       adsurl = {https://ui.adsabs.harvard.edu/abs/2021CQGra..38k5002R}
}

@ARTICLE{rosswog2023,
       author = {{Rosswog}, Stephan and {Torsello}, Francesco and {Diener}, Peter},
        title = "{The Lagrangian Numerical Relativity code SPHINCS\_BSSN\_v1.0}",
      journal = {Front. Appl. Math. Stat.},
         year = 2023,
        month = jun,
       volume = {9},
          eid = {1236586},
        pages = {1236586},
          doi = {10.3389/fams.2023.1236586},
archivePrefix = {arXiv},
       eprint = {2306.06226},
 primaryClass = {gr-qc},
       adsurl = {https://ui.adsabs.harvard.edu/abs/2023arXiv230606226R}
}

@ARTICLE{shankar2026,
       author = {{Shankar}, Swapnil and {Rosswog}, Stephan and {Diener}, Peter},
        title = "{Binary neutron star mergers with tabulated equations of state in SPHINCS\_BSSN}",
      journal = {arXiv e-prints},
         year = 2026,
        month = mar,
          eid = {arXiv:2603.25809},
        pages = {arXiv:2603.25809},
          doi = {10.48550/arXiv.2603.25809},
archivePrefix = {arXiv},
       eprint = {2603.25809},
 primaryClass = {astro-ph.HE},
       adsurl = {https://ui.adsabs.harvard.edu/abs/2026arXiv260325809S}
}

@ARTICLE{biswas2026,
       author = {{Biswas}, Bhaskar and {Rosswog}, Stephan and {Diener}, Peter and {Schnabel}, Lukas},
        title = "{Binary neutron star mergers with SPHINCS\_BSSN: temperature-dependent equations of state and damping of constraint violations}",
      journal = {arXiv e-prints},
         year = 2026,
        month = jan,
          eid = {arXiv:2601.01402},
        pages = {arXiv:2601.01402},
          doi = {10.48550/arXiv.2601.01402},
archivePrefix = {arXiv},
       eprint = {2601.01402},
 primaryClass = {astro-ph.HE},
       adsurl = {https://ui.adsabs.harvard.edu/abs/2026arXiv260101402B}
}

@ARTICLE{mahapatra2026,
       author = {{Mahapatra}, Aryabrat and {Pandey}, Adarsh and {Banerjee}, Pritam and {Sarkar}, Tapobrata},
        title = "{Tidal Disruptions of Close White Dwarf Binaries by Intermediate-mass Black Holes}",
      journal = {\apj},
         year = 2026,
        month = jun,
       volume = {1004},
       number = {2},
          eid = {167},
        pages = {167},
          doi = {10.3847/1538-4357/ae6e3e},
archivePrefix = {arXiv},
       eprint = {2508.03463},
 primaryClass = {astro-ph.HE},
       adsurl = {https://ui.adsabs.harvard.edu/abs/2026ApJ..1004..167M}
}

@ARTICLE{mandel2015,
   author = {{Mandel}, I. and {Levin}, Y.},
    title = "{Double Tidal Disruptions in Galactic Nuclei}",
  journal = {ApJL},
     year = 2015,
    month = "may",
   volume = 805,
      eid = {L4},
    pages = {L4},
      doi = {10.1088/2041-8205/805/1/L4},
   adsurl = {http://adsabs.harvard.edu/abs/2015ApJ...805L...4M},
  archivePrefix = "arXiv",
   eprint = {1504.02787},
}

@book{press1992,
 author = {W. H. Press and B. P. Flannery and S. A. Teukolsky and
  W. T. Vetterling},
 title = {{N}umerical {R}ecipes},
 publisher = {Cambridge University Press},
 year = {1992},
 address = {New York}}

@ARTICLE{rosswog2022,
       author = {{Rosswog}, Stephan and {Diener}, Peter and {Torsello}, Francesco},
        title = "{Thinking Outside the Box: Numerical Relativity with Particles}",
      journal = {Symmetry},
         year = 2022,
        month = jun,
       volume = {14},
       number = {6},
          eid = {1280},
        pages = {1280},
          doi = {10.3390/sym14061280},
archivePrefix = {arXiv},
       eprint = {2205.08130},
 primaryClass = {gr-qc},
       adsurl = {https://ui.adsabs.harvard.edu/abs/2022Symm...14.1280R}
}

@ARTICLE{ramirezruiz09,
       author = {{Ramirez-Ruiz}, Enrico and {Rosswog}, Stephan},
        title = "{The Star Ingesting Luminosity of Intermediate-Mass Black Holes in Globular Clusters}",
      journal = {\apjl},
         year = 2009,
        month = jun,
       volume = {697},
       number = {2},
        pages = {L77-L80},
          doi = {10.1088/0004-637X/697/2/L77},
archivePrefix = {arXiv},
       eprint = {0808.3847},
 primaryClass = {astro-ph},
       adsurl = {https://ui.adsabs.harvard.edu/abs/2009ApJ...697L..77R}
}

@ARTICLE{cocco2026,
       author = {{Cocco}, Marta and {Grignani}, Gianluca and {Harmark}, Troels and {Orselli}, Marta and {Pere{\~n}iguez}, David and {van de Meent}, Maarten},
        title = "{Tidal perturbations of an extreme mass ratio inspiral around a Kerr black hole}",
      journal = {Classical and Quantum Gravity},
         year = 2026,
        month = jul,
       volume = {43},
       number = {14},
          eid = {145005},
        pages = {145005},
          doi = {10.1088/1361-6382/ae80b6},
archivePrefix = {arXiv},
       eprint = {2601.00954},
 primaryClass = {gr-qc},
       adsurl = {https://ui.adsabs.harvard.edu/abs/2026CQGra..43n5005C}
}

@ARTICLE{gottlieb1998,
       author = {{Gottlieb}, S. and {Shu}, C.~W.},
        title = "{Total variation diminishing Runge-Kutta schemes}",
      journal = {Mathematics of Computation},
         year = 1998,
        month = jan,
       volume = {67},
       number = {221},
        pages = {73-85},
          doi = {10.1090/S0025-5718-98-00913-2},
       adsurl = {https://ui.adsabs.harvard.edu/abs/1998MaCom..67...73G}
}

@BOOK{poisson2014,
       author = {{Poisson}, Eric and {Will}, Clifford M.},
        title = "{Gravity}",
         year = 2014,
    publisher = "Cambridge, UK: Cambridge University Press",
       adsurl = {https://ui.adsabs.harvard.edu/abs/2014grav.book.....P}
}

@ARTICLE{gafton2011,
       author = {{Gafton}, Emanuel and {Rosswog}, Stephan},
        title = "{A fast recursive coordinate bisection tree for neighbour search and gravity}",
      journal = {\mnras},
         year = 2011,
        month = dec,
       volume = {418},
       number = {2},
        pages = {770-781},
          doi = {10.1111/j.1365-2966.2011.19528.x},
archivePrefix = {arXiv},
       eprint = {1108.0028},
 primaryClass = {astro-ph.IM},
       adsurl = {https://ui.adsabs.harvard.edu/abs/2011MNRAS.418..770G}
}

@ARTICLE{kato1990,
       author = {{Kato}, Shoji},
        title = "{Trapped One-Armed Corrugation Waves and QPO's}",
      journal = {\pasj},
         year = 1990,
        month = feb,
       volume = {42},
        pages = {99-113},
       adsurl = {https://ui.adsabs.harvard.edu/abs/1990PASJ...42...99K}
}

@ARTICLE{lubow2002,
       author = {{Lubow}, S.~H. and {Ogilvie}, G.~I. and {Pringle}, J.~E.},
        title = "{The evolution of a warped disc around a Kerr black hole}",
      journal = {\mnras},
         year = 2002,
        month = dec,
       volume = {337},
       number = {2},
        pages = {706-712},
          doi = {10.1046/j.1365-8711.2002.05949.x},
archivePrefix = {arXiv},
       eprint = {astro-ph/0208206},
 primaryClass = {astro-ph},
       adsurl = {https://ui.adsabs.harvard.edu/abs/2002MNRAS.337..706L}
}

@ARTICLE{price2018,
       author = {{Price}, Daniel J. and {Wurster}, James and {Tricco}, Terrence S. and {Nixon}, Chris and {Toupin}, St{\'e}ven and {Pettitt}, Alex and {Chan}, Conrad and {Mentiplay}, Daniel and {Laibe}, Guillaume and {Glover}, Simon and {Dobbs}, Clare and {Nealon}, Rebecca and {Liptai}, David and {Worpel}, Hauke and {Bonnerot}, Cl{\'e}ment and {Dipierro}, Giovanni and {Ballabio}, Giulia and {Ragusa}, Enrico and {Federrath}, Christoph and {Iaconi}, Roberto and {Reichardt}, Thomas and {Forgan}, Duncan and {Hutchison}, Mark and {Constantino}, Thomas and {Ayliffe}, Ben and {Hirsh}, Kieran and {Lodato}, Giuseppe},
        title = "{Phantom: A Smoothed Particle Hydrodynamics and Magnetohydrodynamics Code for Astrophysics}",
      journal = {\pasa},
         year = 2018,
        month = sep,
       volume = {35},
          eid = {e031},
        pages = {e031},
          doi = {10.1017/pasa.2018.25},
archivePrefix = {arXiv},
       eprint = {1702.03930},
 primaryClass = {astro-ph.IM},
       adsurl = {https://ui.adsabs.harvard.edu/abs/2018PASA...35...31P}
}

@ARTICLE{hopkins2015,
       author = {{Hopkins}, Philip F.},
        title = "{A new class of accurate, mesh-free hydrodynamic simulation methods}",
      journal = {\mnras},
         year = 2015,
        month = jun,
       volume = {450},
       number = {1},
        pages = {53-110},
          doi = {10.1093/mnras/stv195},
archivePrefix = {arXiv},
       eprint = {1409.7395},
 primaryClass = {astro-ph.CO},
       adsurl = {https://ui.adsabs.harvard.edu/abs/2015MNRAS.450...53H}
}

@ARTICLE{michel1972,
       author = {{Michel}, F. Curtis},
        title = "{Accretion of Matter by Condensed Objects}",
      journal = {\apss},
         year = 1972,
        month = jan,
       volume = {15},
       number = {1},
        pages = {153-160},
          doi = {10.1007/BF00649949},
       adsurl = {https://ui.adsabs.harvard.edu/abs/1972Ap&SS..15..153M}
}

@ARTICLE{hawley1984,
       author = {{Hawley}, J.~F. and {Smarr}, L.~L. and {Wilson}, J.~R.},
        title = "{A numerical study of nonspherical black hole accretion. I Equations and test problems}",
      journal = {\apj},
         year = 1984,
        month = feb,
       volume = {277},
        pages = {296-311},
          doi = {10.1086/161696},
       adsurl = {https://ui.adsabs.harvard.edu/abs/1984ApJ...277..296H}
}

@ARTICLE{rosswog2015a,
       author = {{Rosswog}, S.},
        title = "{Boosting the accuracy of SPH techniques: Newtonian and special-relativistic tests}",
      journal = {\mnras},
         year = 2015,
        month = apr,
       volume = {448},
       number = {4},
        pages = {3628-3664},
          doi = {10.1093/mnras/stv225},
archivePrefix = {arXiv},
       eprint = {1405.6034},
 primaryClass = {astro-ph.IM},
       adsurl = {https://ui.adsabs.harvard.edu/abs/2015MNRAS.448.3628R}
}

@article{wendland1995,
	author = {Wendland, Holger},
	date = {1995/12/01},
	doi = {10.1007/BF02123482},
	id = {Wendland1995},
	isbn = {1572-9044},
	journal = {Advances in Computational Mathematics},
	number = {1},
	pages = {389--396},
	title = {Piecewise polynomial, positive definite and compactly supported radial functions of minimal degree},
	url = {https://doi.org/10.1007/BF02123482},
	volume = {4},
	year = {1995}
}

@ARTICLE{rosswog2020,
       author = {{Rosswog}, S.},
        title = "{The Lagrangian hydrodynamics code MAGMA2}",
      journal = {\mnras},
         year = 2020,
        month = nov,
       volume = {498},
       number = {3},
        pages = {4230-4255},
          doi = {10.1093/mnras/staa2591},
archivePrefix = {arXiv},
       eprint = {1911.13093},
 primaryClass = {astro-ph.IM},
       adsurl = {https://ui.adsabs.harvard.edu/abs/2020MNRAS.498.4230R}
}

@ARTICLE{rosswog2007,
       author = {{Rosswog}, Stephan and {Price}, Daniel},
        title = "{MAGMA: a three-dimensional, Lagrangian magnetohydrodynamics code for merger applications}",
      journal = {\mnras},
         year = 2007,
        month = aug,
       volume = {379},
       number = {3},
        pages = {915-931},
          doi = {10.1111/j.1365-2966.2007.11984.x},
archivePrefix = {arXiv},
       eprint = {0705.1441},
 primaryClass = {astro-ph},
       adsurl = {https://ui.adsabs.harvard.edu/abs/2007MNRAS.379..915R}
}

@ARTICLE{yalinewich2015,
       author = {{Yalinewich}, Almog and {Steinberg}, Elad and {Sari}, Re'em},
        title = "{RICH: Open-source Hydrodynamic Simulation on a Moving Voronoi Mesh}",
      journal = {\apjs},
         year = 2015,
        month = feb,
       volume = {216},
       number = {2},
          eid = {35},
        pages = {35},
          doi = {10.1088/0067-0049/216/2/35},
archivePrefix = {arXiv},
       eprint = {1410.3219},
 primaryClass = {astro-ph.IM},
       adsurl = {https://ui.adsabs.harvard.edu/abs/2015ApJS..216...35Y}
}

@ARTICLE{springel2010,
       author = {{Springel}, Volker},
        title = "{E pur si muove: Galilean-invariant cosmological hydrodynamical simulations on a moving mesh}",
      journal = {\mnras},
         year = 2010,
        month = jan,
       volume = {401},
       number = {2},
        pages = {791-851},
          doi = {10.1111/j.1365-2966.2009.15715.x},
archivePrefix = {arXiv},
       eprint = {0901.4107},
 primaryClass = {astro-ph.CO},
       adsurl = {https://ui.adsabs.harvard.edu/abs/2010MNRAS.401..791S}
}

@ARTICLE{springel2021,
       author = {{Springel}, Volker and {Pakmor}, R{\"u}diger and {Zier}, Oliver and {Reinecke}, Martin},
        title = "{Simulating cosmic structure formation with the GADGET-4 code}",
      journal = {\mnras},
         year = 2021,
        month = sep,
       volume = {506},
       number = {2},
        pages = {2871-2949},
          doi = {10.1093/mnras/stab1855},
archivePrefix = {arXiv},
       eprint = {2010.03567},
 primaryClass = {astro-ph.IM},
       adsurl = {https://ui.adsabs.harvard.edu/abs/2021MNRAS.506.2871S}
}

@ARTICLE{steinberg2024,
       author = {{Steinberg}, Elad and {Stone}, Nicholas C.},
        title = "{Stream-disk shocks as the origins of peak light in tidal disruption events}",
      journal = {\nat},
         year = 2024,
        month = jan,
       volume = {625},
       number = {7995},
        pages = {463-467},
          doi = {10.1038/s41586-023-06875-y},
archivePrefix = {arXiv},
       eprint = {2206.10641},
 primaryClass = {astro-ph.HE},
       adsurl = {https://ui.adsabs.harvard.edu/abs/2024Natur.625..463S}
}

@ARTICLE{price2024,
       author = {{Price}, Daniel J. and {Liptai}, David and {Mandel}, Ilya and {Shepherd}, Joanna and {Lodato}, Giuseppe and {Levin}, Yuri},
        title = "{Eddington Envelopes: The Fate of Stars on Parabolic Orbits Tidally Disrupted by Supermassive Black Holes}",
      journal = {\apjl},
         year = 2024,
        month = aug,
       volume = {971},
       number = {2},
          eid = {L46},
        pages = {L46},
          doi = {10.3847/2041-8213/ad6862},
archivePrefix = {arXiv},
       eprint = {2404.09381},
 primaryClass = {astro-ph.HE},
       adsurl = {https://ui.adsabs.harvard.edu/abs/2024ApJ...971L..46P}
}

@ARTICLE{goicovic2019,
       author = {{Goicovic}, Felipe G. and {Springel}, Volker and {Ohlmann}, Sebastian T. and {Pakmor}, R{\"u}diger},
        title = "{Hydrodynamical moving-mesh simulations of the tidal disruption of stars by supermassive black holes}",
      journal = {\mnras},
         year = 2019,
        month = jul,
       volume = {487},
       number = {1},
        pages = {981-992},
          doi = {10.1093/mnras/stz1368},
archivePrefix = {arXiv},
       eprint = {1902.08202},
 primaryClass = {astro-ph.HE},
       adsurl = {https://ui.adsabs.harvard.edu/abs/2019MNRAS.487..981G}
}

@ARTICLE{bonnerot2017,
       author = {{Bonnerot}, Cl{\'e}ment and {Price}, Daniel J. and {Lodato}, Giuseppe and {Rossi}, Elena M.},
        title = "{Magnetic field evolution in tidal disruption events}",
      journal = {\mnras},
         year = 2017,
        month = aug,
       volume = {469},
       number = {4},
        pages = {4879-4888},
          doi = {10.1093/mnras/stx1210},
archivePrefix = {arXiv},
       eprint = {1611.09853},
 primaryClass = {astro-ph.HE},
       adsurl = {https://ui.adsabs.harvard.edu/abs/2017MNRAS.469.4879B}
}

@phdthesis{tejeda2012,
  author       = {Tejeda, Emilio}, 
  title        = {An analytic Kerr-accretion model as a test solution for a new GR SPH code},
  school       = {International School for Advanced Studies of
Trieste, Italy},
  year         = 2012,
}

@ARTICLE{springel2002,
       author = {{Springel}, Volker and {Hernquist}, Lars},
        title = "{Cosmological smoothed particle hydrodynamics simulations: the entropy equation}",
      journal = {\mnras},
         year = 2002,
        month = jul,
       volume = {333},
       number = {3},
        pages = {649-664},
          doi = {10.1046/j.1365-8711.2002.05445.x},
archivePrefix = {arXiv},
       eprint = {astro-ph/0111016},
 primaryClass = {astro-ph},
       adsurl = {https://ui.adsabs.harvard.edu/abs/2002MNRAS.333..649S}
}

@ARTICLE{calderon2024,
       author = {{Calder{\'o}n}, Diego and {Pejcha}, Ond{\v{r}}ej and {Metzger}, Brian D. and {Duffell}, Paul C.},
        title = "{The effect of relativistic precession on light curves of tidal disruption events}",
      journal = {\mnras},
         year = 2024,
        month = feb,
       volume = {528},
       number = {2},
        pages = {2568-2587},
          doi = {10.1093/mnras/stae194},
archivePrefix = {arXiv},
       eprint = {2309.10040},
 primaryClass = {astro-ph.HE},
       adsurl = {https://ui.adsabs.harvard.edu/abs/2024MNRAS.528.2568C}
}

@ARTICLE{tejeda2017,
       author = {{Tejeda}, Emilio and {Gafton}, Emanuel and {Rosswog}, Stephan and {Miller}, John C.},
        title = "{Tidal disruptions by rotating black holes: relativistic hydrodynamics with Newtonian codes}",
      journal = {\mnras},
         year = 2017,
        month = aug,
       volume = {469},
       number = {4},
        pages = {4483-4503},
          doi = {10.1093/mnras/stx1089},
archivePrefix = {arXiv},
       eprint = {1701.00303},
 primaryClass = {astro-ph.HE},
       adsurl = {https://ui.adsabs.harvard.edu/abs/2017MNRAS.469.4483T}
}

@ARTICLE{tejeda2013,
       author = {{Tejeda}, Emilio and {Rosswog}, Stephan},
        title = "{An accurate Newtonian description of particle motion around a Schwarzschild black hole}",
      journal = {\mnras},
         year = 2013,
        month = aug,
       volume = {433},
       number = {3},
        pages = {1930-1940},
          doi = {10.1093/mnras/stt853},
archivePrefix = {arXiv},
       eprint = {1303.4068},
 primaryClass = {astro-ph.HE},
       adsurl = {https://ui.adsabs.harvard.edu/abs/2013MNRAS.433.1930T}
}

@ARTICLE{pakmor2016,
       author = {{Pakmor}, R{\"u}diger and {Springel}, Volker and {Bauer}, Andreas and {Mocz}, Philip and {Munoz}, Diego J. and {Ohlmann}, Sebastian T. and {Schaal}, Kevin and {Zhu}, Chenchong},
        title = "{Improving the convergence properties of the moving-mesh code AREPO}",
      journal = {\mnras},
         year = 2016,
        month = jan,
       volume = {455},
       number = {1},
        pages = {1134-1143},
          doi = {10.1093/mnras/stv2380},
archivePrefix = {arXiv},
       eprint = {1503.00562},
 primaryClass = {astro-ph.GA},
       adsurl = {https://ui.adsabs.harvard.edu/abs/2016MNRAS.455.1134P}
}

@ARTICLE{gafton2019,
       author = {{Gafton}, Emanuel and {Rosswog}, Stephan},
        title = "{Tidal disruptions by rotating black holes: effects of spin and impact parameter}",
      journal = {\mnras},
         year = 2019,
        month = aug,
       volume = {487},
       number = {4},
        pages = {4790-4808},
          doi = {10.1093/mnras/stz1530},
archivePrefix = {arXiv},
       eprint = {1903.09147},
 primaryClass = {astro-ph.HE},
       adsurl = {https://ui.adsabs.harvard.edu/abs/2019MNRAS.487.4790G}
}

@ARTICLE{jankovic2023,
       author = {{Jankovi{\v{c}}}, T. and {Gomboc}, A.},
        title = "{The Mass Fallback Rate of the Debris in Relativistic Stellar Tidal Disruption Events}",
      journal = {\apj},
         year = 2023,
        month = mar,
       volume = {946},
       number = {1},
          eid = {25},
        pages = {25},
          doi = {10.3847/1538-4357/acb8b0},
archivePrefix = {arXiv},
       eprint = {2302.00607},
 primaryClass = {astro-ph.HE},
       adsurl = {https://ui.adsabs.harvard.edu/abs/2023ApJ...946...25J}
}

@ARTICLE{hu2026,
       author = {{Hu}, Fangyi (Fitz) and {Mandel}, Ilya and {Nealon}, Rebecca and {Price}, Daniel J.},
        title = "{Converged Simulations of the Nozzle Shock in Tidal Disruption Events}",
      journal = {\apjl},
         year = 2026,
        month = jan,
       volume = {996},
       number = {2},
          eid = {L21},
        pages = {L21},
          doi = {10.3847/2041-8213/ae27cc},
archivePrefix = {arXiv},
       eprint = {2510.04790},
 primaryClass = {astro-ph.HE},
       adsurl = {https://ui.adsabs.harvard.edu/abs/2026ApJ...996L..21H}
}

@ARTICLE{kubli2026,
       author = {{Kubli}, Noah and {Franchini}, Alessia and {Coughlin}, Eric R. and {Nixon}, C.~J. and {Keller}, Sebastian and {Capelo}, Pedro R. and {Mayer}, Lucio},
        title = "{Tidal Disruption Events with SPH-EXA: Resolving the Return of the Stream}",
      journal = {\apjl},
         year = 2026,
        month = mar,
       volume = {999},
       number = {2},
          eid = {L40},
        pages = {L40},
          doi = {10.3847/2041-8213/ae4748},
archivePrefix = {arXiv},
       eprint = {2510.26663},
 primaryClass = {astro-ph.HE},
       adsurl = {https://ui.adsabs.harvard.edu/abs/2026ApJ...999L..40K}
}

@ARTICLE{calderon2026,
       author = {{Calder{\'o}n}, Diego and {Pejcha}, Ond{\v{r}}ej and {Metzger}, Brian D. and {Duffell}, Paul C. and {Rosswog}, Stephan},
        title = "{Quantifying the impact of relativistic precession on tidal disruption event light curves}",
      journal = {Astronomische Nachricthen},
         year = 2026,
        month = mar,
          eid = {e70092},
        pages = {e70092},
          doi = {10.1002/asna.70092},
archivePrefix = {arXiv},
       eprint = {2603.10124},
 primaryClass = {astro-ph.HE},
       adsurl = {https://ui.adsabs.harvard.edu/abs/2026arXiv260310124C}
}

@ARTICLE{abramowicz1978,
       author = {{Abramowicz}, M. and {Jaroszynski}, M. and {Sikora}, M.},
        title = "{Relativistic, accreting disks.}",
      journal = {\aap},
         year = 1978,
        month = feb,
       volume = {63},
        pages = {221-224},
       adsurl = {https://ui.adsabs.harvard.edu/abs/1978A&A....63..221A}
}

@ARTICLE{chan2026,
       author = {{Chan}, Ho-Sang and {Ryu}, Taeho and {Krolik}, Julian and {Piran}, Tsvi},
        title = "{Unexpectedly Weak General Relativistic Effects in Strongly Relativistic Tidal Disruption Events}",
      journal = {\apj},
         year = 2026,
        month = jun,
       volume = {1003},
       number = {2},
          eid = {221},
        pages = {221},
          doi = {10.3847/1538-4357/ae67fa},
archivePrefix = {arXiv},
       eprint = {2603.10208},
 primaryClass = {astro-ph.HE},
       adsurl = {https://ui.adsabs.harvard.edu/abs/2026ApJ..1003..221C}
}

@ARTICLE{cheng2014,
       author = {{Cheng}, Roseanne M. and {Bogdanovi{\'c}}, Tamara},
        title = "{Tidal disruption of a star in the Schwarzschild spacetime: Relativistic effects in the return rate of debris}",
      journal = {\prd},
         year = 2014,
        month = sep,
       volume = {90},
       number = {6},
          eid = {064020},
        pages = {064020},
          doi = {10.1103/PhysRevD.90.064020},
archivePrefix = {arXiv},
       eprint = {1407.3266},
 primaryClass = {gr-qc},
       adsurl = {https://ui.adsabs.harvard.edu/abs/2014PhRvD..90f4020C}
}

@ARTICLE{virtanen2020,
  author  = {Virtanen, Pauli and Gommers, Ralf and Oliphant, Travis E.  and
            Haberland, Matt and Reddy, Tyler and Cournapeau, David and
            Burovski, Evgeni and Peterson, Pearu and Weckesser, Warren and
            Bright, Jonathan and {van der Walt}, St{\'e}fan J. and
            Brett, Matthew and Wilson, Joshua and Millman, K. Jarrod and
            Mayorov, Nikolay and Nelson, Andrew R. J. and Jones, Eric and
            Kern, Robert and Larson, Eric and Carey, C J and
            Polat, {\.I}lhan and Feng, Yu and Moore, Eric W. and
            {VanderPlas}, Jake and Laxalde, Denis and Perktold, Josef and
            Cimrman, Robert and Henriksen, Ian and Quintero, E. A. and
            Harris, Charles R. and Archibald, Anne M. and
            Ribeiro, Ant{\^o}nio H. and Pedregosa, Fabian and
            {van Mulbregt}, Paul and {SciPy 1.0 Contributors}},
  title   = {{{SciPy} 1.0: Fundamental Algorithms for Scientific
            Computing in Python}},
  journal = {Nature Methods},
  year    = {2020},
  volume  = {17},
  pages   = {261--272},
  adsurl  = {https://rdcu.be/b08Wh},
  doi     = {10.1038/s41592-019-0686-2},
}

@ARTICLE{evans1989,
       author = {{Evans}, Charles R. and {Kochanek}, Christopher S.},
        title = "{The Tidal Disruption of a Star by a Massive Black Hole}",
      journal = {\apjl},
         year = 1989,
        month = nov,
       volume = {346},
        pages = {L13},
          doi = {10.1086/185567},
       adsurl = {https://ui.adsabs.harvard.edu/abs/1989ApJ...346L..13E}
}

@ARTICLE{hills1975,
       author = {{Hills}, J.~G.},
        title = "{Possible power source of Seyfert galaxies and QSOs}",
      journal = {\nat},
         year = 1975,
        month = mar,
       volume = {254},
       number = {5498},
        pages = {295-298},
          doi = {10.1038/254295a0},
       adsurl = {https://ui.adsabs.harvard.edu/abs/1975Natur.254..295H}
}

@ARTICLE{rees1988,
       author = {{Rees}, Martin J.},
        title = "{Tidal disruption of stars by black holes of {}10$^{6}$-{}10$^{8}$ solar masses in nearby galaxies}",
      journal = {\nat},
         year = 1988,
        month = jun,
       volume = {333},
       number = {6173},
        pages = {523-528},
          doi = {10.1038/333523a0},
       adsurl = {https://ui.adsabs.harvard.edu/abs/1988Natur.333..523R}
}

@ARTICLE{alexander2005,
       author = {{Alexander}, Tal},
        title = "{Stellar processes near the massive black hole in the Galactic center [review article]}",
      journal = {\physrep},
         year = 2005,
        month = nov,
       volume = {419},
       number = {2-3},
        pages = {65-142},
          doi = {10.1016/j.physrep.2005.08.002},
archivePrefix = {arXiv},
       eprint = {astro-ph/0508106},
 primaryClass = {astro-ph},
       adsurl = {https://ui.adsabs.harvard.edu/abs/2005PhR...419...65A}
}

@ARTICLE{gezari2021,
       author = {{Gezari}, Suvi},
        title = "{Tidal Disruption Events}",
      journal = {\araa},
         year = 2021,
        month = sep,
       volume = {59},
        pages = {21-58},
          doi = {10.1146/annurev-astro-111720-030029},
archivePrefix = {arXiv},
       eprint = {2104.14580},
 primaryClass = {astro-ph.HE},
       adsurl = {https://ui.adsabs.harvard.edu/abs/2021ARA&A..59...21G}
}

@INPROCEEDINGS{phinney1989,
       author = {{Phinney}, E.~S.},
        title = "{Manifestations of a Massive Black Hole in the Galactic Center}",
    booktitle = {The Center of the Galaxy},
         year = 1989,
       editor = {{Morris}, Mark},
       series = {IAU Symposium},
       volume = {136},
        month = jan,
        pages = {543},
       adsurl = {https://ui.adsabs.harvard.edu/abs/1989IAUS..136..543P}
}

@ARTICLE{bonnerot2020,
       author = {{Bonnerot}, Cl{\'e}ment and {Lu}, Wenbin},
        title = "{Simulating disc formation in tidal disruption events}",
      journal = {\mnras},
         year = 2020,
        month = jun,
       volume = {495},
       number = {1},
        pages = {1374-1391},
          doi = {10.1093/mnras/staa1246},
archivePrefix = {arXiv},
       eprint = {1906.05865},
 primaryClass = {astro-ph.HE},
       adsurl = {https://ui.adsabs.harvard.edu/abs/2020MNRAS.495.1374B}
}

@ARTICLE{piran2015,
       author = {{Piran}, Tsvi and {Svirski}, Gilad and {Krolik}, Julian and {Cheng}, Roseanne M. and {Shiokawa}, Hotaka},
        title = "{′Disk Formation Versus Disk Accretion{\textemdash}What Powers Tidal Disruption Events?}",
      journal = {\apj},
         year = 2015,
        month = jun,
       volume = {806},
       number = {2},
          eid = {164},
        pages = {164},
          doi = {10.1088/0004-637X/806/2/164},
archivePrefix = {arXiv},
       eprint = {1502.05792},
 primaryClass = {astro-ph.HE},
       adsurl = {https://ui.adsabs.harvard.edu/abs/2015ApJ...806..164P}
}

@ARTICLE{jiang2016,
       author = {{Jiang}, Ning and {Dou}, Liming and {Wang}, Tinggui and {Yang}, Chenwei and {Lyu}, Jianwei and {Zhou}, Hongyan},
        title = "{The WISE Detection of an Infrared Echo in Tidal Disruption Event ASASSN-14li}",
      journal = {\apjl},
         year = 2016,
        month = sep,
       volume = {828},
       number = {1},
          eid = {L14},
        pages = {L14},
          doi = {10.3847/2041-8205/828/1/L14},
archivePrefix = {arXiv},
       eprint = {1605.04640},
 primaryClass = {astro-ph.HE},
       adsurl = {https://ui.adsabs.harvard.edu/abs/2016ApJ...828L..14J}
}

@ARTICLE{metzger2016,
       author = {{Metzger}, Brian D. and {Stone}, Nicholas C.},
        title = "{A bright year for tidal disruptions}",
      journal = {\mnras},
         year = 2016,
        month = sep,
       volume = {461},
       number = {1},
        pages = {948-966},
          doi = {10.1093/mnras/stw1394},
archivePrefix = {arXiv},
       eprint = {1506.03453},
 primaryClass = {astro-ph.HE},
       adsurl = {https://ui.adsabs.harvard.edu/abs/2016MNRAS.461..948M}
}

@ARTICLE{ryu2023,
       author = {{Ryu}, Taeho and {Krolik}, Julian and {Piran}, Tsvi and {Noble}, Scott C. and {Avara}, Mark},
        title = "{Shocks Power Tidal Disruption Events}",
      journal = {\apj},
         year = 2023,
        month = nov,
       volume = {957},
       number = {1},
          eid = {12},
        pages = {12},
          doi = {10.3847/1538-4357/acf5de},
archivePrefix = {arXiv},
       eprint = {2305.05333},
 primaryClass = {astro-ph.HE},
       adsurl = {https://ui.adsabs.harvard.edu/abs/2023ApJ...957...12R}
}

@ARTICLE{langis2026,
       author = {{Langis}, D.~A. and {Liodakis}, I. and {Koljonen}, K.~I.~I. and {Paggi}, A. and {Globus}, N. and {Wyrzykowski}, L. and {Miko{\l}ajczyk}, P.~J. and {Kotysz}, K. and {Zieli{\'n}ski}, P. and {Ihanec}, N. and {Ding}, J. and {Morshed}, D. and {Torres}, Z.},
        title = "{Repeating flares, X-ray outbursts and delayed infrared emission: A comprehensive compilation of optical tidal disruption events: TDECat}",
      journal = {\aap},
         year = 2026,
        month = mar,
       volume = {707},
          eid = {A171},
        pages = {A171},
          doi = {10.1051/0004-6361/202554882},
archivePrefix = {arXiv},
       eprint = {2506.05476},
 primaryClass = {astro-ph.HE},
       adsurl = {https://ui.adsabs.harvard.edu/abs/2026A&A...707A.171L}
}

@ARTICLE{hammerstein2023,
       author = {{Hammerstein}, Erica and {van Velzen}, Sjoert and {Gezari}, Suvi and {Cenko}, S. Bradley and {Yao}, Yuhan and {Ward}, Charlotte and {Frederick}, Sara and {Villanueva}, Natalia and {Somalwar}, Jean J. and {Graham}, Matthew J. and {Kulkarni}, Shrinivas R. and {Stern}, Daniel and {Andreoni}, Igor and {Bellm}, Eric C. and {Dekany}, Richard and {Dhawan}, Suhail and {Drake}, Andrew J. and {Fremling}, Christoffer and {Gatkine}, Pradip and {Groom}, Steven L. and {Ho}, Anna Y.~Q. and {Kasliwal}, Mansi M. and {Karambelkar}, Viraj and {Kool}, Erik C. and {Masci}, Frank J. and {Medford}, Michael S. and {Perley}, Daniel A. and {Purdum}, Josiah and {van Roestel}, Jan and {Sharma}, Yashvi and {Sollerman}, Jesper and {Taggart}, Kirsty and {Yan}, Lin},
        title = "{The Final Season Reimagined: 30 Tidal Disruption Events from the ZTF-I Survey}",
      journal = {\apj},
         year = 2023,
        month = jan,
       volume = {942},
       number = {1},
          eid = {9},
        pages = {9},
          doi = {10.3847/1538-4357/aca283},
archivePrefix = {arXiv},
       eprint = {2203.01461},
 primaryClass = {astro-ph.HE},
       adsurl = {https://ui.adsabs.harvard.edu/abs/2023ApJ...942....9H}
}

@ARTICLE{vanvelzen2020,
       author = {{van Velzen}, Sjoert and {Holoien}, Thomas W.-S. and {Onori}, Francesca and {Hung}, Tiara and {Arcavi}, Iair},
        title = "{Optical-Ultraviolet Tidal Disruption Events}",
      journal = {\ssr},
         year = 2020,
        month = oct,
       volume = {216},
       number = {8},
          eid = {124},
        pages = {124},
          doi = {10.1007/s11214-020-00753-z},
archivePrefix = {arXiv},
       eprint = {2008.05461},
 primaryClass = {astro-ph.HE},
       adsurl = {https://ui.adsabs.harvard.edu/abs/2020SSRv..216..124V}
}

@ARTICLE{pacuraru2026,
       author = {{Pacuraru}, Simona and {Bonnerot}, Cl{\'e}ment and {Pessah}, Martin E.},
        title = "{The impact of magnetic fields during tidal disruption events}",
      journal = {\mnras},
         year = 2026,
        month = may,
       volume = {548},
       number = {2},
          eid = {stag603},
        pages = {stag603},
          doi = {10.1093/mnras/stag603},
archivePrefix = {arXiv},
       eprint = {2511.21818},
 primaryClass = {astro-ph.HE},
       adsurl = {https://ui.adsabs.harvard.edu/abs/2026MNRAS.548ag603P}
}

@ARTICLE{abolmasov2026,
       author = {{Abolmasov}, Pavel and {Bromberg}, Omer and {Levinson}, Amir and {Nakar}, Ehud},
        title = "{Tidal Disruption of a Magnetized Star}",
      journal = {\apj},
         year = 2026,
        month = apr,
       volume = {1001},
       number = {1},
          eid = {71},
        pages = {71},
          doi = {10.3847/1538-4357/ae4ebc},
archivePrefix = {arXiv},
       eprint = {2509.23894},
 primaryClass = {astro-ph.HE},
       adsurl = {https://ui.adsabs.harvard.edu/abs/2026ApJ..1001...71A}
}

@ARTICLE{dai2018,
       author = {{Dai}, Lixin and {McKinney}, Jonathan C. and {Roth}, Nathaniel and {Ramirez-Ruiz}, Enrico and {Miller}, M. Coleman},
        title = "{A Unified Model for Tidal Disruption Events}",
      journal = {\apjl},
         year = 2018,
        month = jun,
       volume = {859},
       number = {2},
          eid = {L20},
        pages = {L20},
          doi = {10.3847/2041-8213/aab429},
archivePrefix = {arXiv},
       eprint = {1803.03265},
 primaryClass = {astro-ph.HE},
       adsurl = {https://ui.adsabs.harvard.edu/abs/2018ApJ...859L..20D}
}

@ARTICLE{stone2020,
       author = {{Stone}, N.~C. and {Vasiliev}, E. and {Kesden}, M. and {Rossi}, E.~M. and {Perets}, H.~B. and {Amaro-Seoane}, P.},
        title = "{Rates of Stellar Tidal Disruption}",
      journal = {\ssr},
         year = 2020,
        month = mar,
       volume = {216},
       number = {3},
          eid = {35},
        pages = {35},
          doi = {10.1007/s11214-020-00651-4},
archivePrefix = {arXiv},
       eprint = {2003.08953},
 primaryClass = {astro-ph.HE},
       adsurl = {https://ui.adsabs.harvard.edu/abs/2020SSRv..216...35S}
}

@ARTICLE{ivezic2019,
       author = {{Ivezi{\'c}}, {\v{Z}}eljko and {Kahn}, Steven M. and {Tyson}, J. Anthony and {Abel}, Bob and {Acosta}, Emily and {Allsman}, Robyn and {Alonso}, David and {AlSayyad}, Yusra and {Anderson}, Scott F. and {Andrew}, John and {Angel}, James Roger P. and {Angeli}, George Z. and {Ansari}, Reza and {Antilogus}, Pierre and {Araujo}, Constanza and {Armstrong}, Robert and {Arndt}, Kirk T. and {Astier}, Pierre and {Aubourg}, {\'E}ric and {Auza}, Nicole and {Axelrod}, Tim S. and {Bard}, Deborah J. and {Barr}, Jeff D. and {Barrau}, Aurelian and {Bartlett}, James G. and {Bauer}, Amanda E. and {Bauman}, Brian J. and {Baumont}, Sylvain and {Bechtol}, Ellen and {Bechtol}, Keith and {Becker}, Andrew C. and {Becla}, Jacek and {Beldica}, Cristina and {Bellavia}, Steve and {Bianco}, Federica B. and {Biswas}, Rahul and {Blanc}, Guillaume and {Blazek}, Jonathan and {Blandford}, Roger D. and {Bloom}, Josh S. and {Bogart}, Joanne and {Bond}, Tim W. and {Booth}, Michael T. and {Borgland}, Anders W. and {Borne}, Kirk and {Bosch}, James F. and {Boutigny}, Dominique and {Brackett}, Craig A. and {Bradshaw}, Andrew and {Brandt}, William Nielsen and {Brown}, Michael E. and {Bullock}, James S. and {Burchat}, Patricia and {Burke}, David L. and {Cagnoli}, Gianpietro and {Calabrese}, Daniel and {Callahan}, Shawn and {Callen}, Alice L. and {Carlin}, Jeffrey L. and {Carlson}, Erin L. and {Chandrasekharan}, Srinivasan and {Charles-Emerson}, Glenaver and {Chesley}, Steve and {Cheu}, Elliott C. and {Chiang}, Hsin-Fang and {Chiang}, James and {Chirino}, Carol and {Chow}, Derek and {Ciardi}, David R. and {Claver}, Charles F. and {Cohen-Tanugi}, Johann and {Cockrum}, Joseph J. and {Coles}, Rebecca and {Connolly}, Andrew J. and {Cook}, Kem H. and {Cooray}, Asantha and {Covey}, Kevin R. and {Cribbs}, Chris and {Cui}, Wei and {Cutri}, Roc and {Daly}, Philip N. and {Daniel}, Scott F. and {Daruich}, Felipe and {Daubard}, Guillaume and {Daues}, Greg and {Dawson}, William and {Delgado}, Francisco and {Dellapenna}, Alfred and {de Peyster}, Robert and {de Val-Borro}, Miguel and {Digel}, Seth W. and {Doherty}, Peter and {Dubois}, Richard and {Dubois-Felsmann}, Gregory P. and {Durech}, Josef and {Economou}, Frossie and {Eifler}, Tim and {Eracleous}, Michael and {Emmons}, Benjamin L. and {Fausti Neto}, Angelo and {Ferguson}, Henry and {Figueroa}, Enrique and {Fisher-Levine}, Merlin and {Focke}, Warren and {Foss}, Michael D. and {Frank}, James and {Freemon}, Michael D. and {Gangler}, Emmanuel and {Gawiser}, Eric and {Geary}, John C. and {Gee}, Perry and {Geha}, Marla and {Gessner}, Charles J.~B. and {Gibson}, Robert R. and {Gilmore}, D. Kirk and {Glanzman}, Thomas and {Glick}, William and {Goldina}, Tatiana and {Goldstein}, Daniel A. and {Goodenow}, Iain and {Graham}, Melissa L. and {Gressler}, William J. and {Gris}, Philippe and {Guy}, Leanne P. and {Guyonnet}, Augustin and {Haller}, Gunther and {Harris}, Ron and {Hascall}, Patrick A. and {Haupt}, Justine and {Hernandez}, Fabio and {Herrmann}, Sven and {Hileman}, Edward and {Hoblitt}, Joshua and {Hodgson}, John A. and {Hogan}, Craig and {Howard}, James D. and {Huang}, Dajun and {Huffer}, Michael E. and {Ingraham}, Patrick and {Innes}, Walter R. and {Jacoby}, Suzanne H. and {Jain}, Bhuvnesh and {Jammes}, Fabrice and {Jee}, M. James and {Jenness}, Tim and {Jernigan}, Garrett and {Jevremovi{\'c}}, Darko and {Johns}, Kenneth and {Johnson}, Anthony S. and {Johnson}, Margaret W.~G. and {Jones}, R. Lynne and {Juramy-Gilles}, Claire and {Juri{\'c}}, Mario and {Kalirai}, Jason S. and {Kallivayalil}, Nitya J. and {Kalmbach}, Bryce and {Kantor}, Jeffrey P. and {Karst}, Pierre and {Kasliwal}, Mansi M. and {Kelly}, Heather and {Kessler}, Richard and {Kinnison}, Veronica and {Kirkby}, David and {Knox}, Lloyd and {Kotov}, Ivan V. and {Krabbendam}, Victor L. and {Krughoff}, K. Simon and {Kub{\'a}nek}, Petr and {Kuczewski}, John and {Kulkarni}, Shri and {Ku}, John and {Kurita}, Nadine R. and {Lage}, Craig S. and {Lambert}, Ron and {Lange}, Travis and {Langton}, J. Brian and {Le Guillou}, Laurent and {Levine}, Deborah and {Liang}, Ming and {Lim}, Kian-Tat and {Lintott}, Chris J. and {Long}, Kevin E. and {Lopez}, Margaux and {Lotz}, Paul J. and {Lupton}, Robert H. and {Lust}, Nate B. and {MacArthur}, Lauren A. and {Mahabal}, Ashish and {Mandelbaum}, Rachel and {Markiewicz}, Thomas W. and {Marsh}, Darren S. and {Marshall}, Philip J. and {Marshall}, Stuart and {May}, Morgan and {McKercher}, Robert and {McQueen}, Michelle and {Meyers}, Joshua and {Migliore}, Myriam and {Miller}, Michelle and {Mills}, David J.},
        title = "{LSST: From Science Drivers to Reference Design and Anticipated Data Products}",
      journal = {\apj},
         year = 2019,
        month = mar,
       volume = {873},
       number = {2},
          eid = {111},
        pages = {111},
          doi = {10.3847/1538-4357/ab042c},
archivePrefix = {arXiv},
       eprint = {0805.2366},
 primaryClass = {astro-ph},
       adsurl = {https://ui.adsabs.harvard.edu/abs/2019ApJ...873..111I}
}

@ARTICLE{amaro2025,
       author = {{Amaro Seoane}, Pau},
        title = "{Illuminating gravitational wave sources with Sgr A* flares}",
      journal = {arXiv e-prints},
         year = 2025,
        month = oct,
          eid = {arXiv:2510.20898},
        pages = {arXiv:2510.20898},
          doi = {10.48550/arXiv.2510.20898},
archivePrefix = {arXiv},
       eprint = {2510.20898},
 primaryClass = {astro-ph.HE},
       adsurl = {https://ui.adsabs.harvard.edu/abs/2025arXiv251020898A}
}

@ARTICLE{liptai2019b,
       author = {{Liptai}, David and {Price}, Daniel J. and {Mandel}, Ilya and {Lodato}, Giuseppe},
        title = "{Disc formation from tidal disruption of stars on eccentric orbits by Kerr black holes using GRSPH}",
      journal = {arXiv e-prints},
         year = 2019,
        month = oct,
          eid = {arXiv:1910.10154},
        pages = {arXiv:1910.10154},
          doi = {10.48550/arXiv.1910.10154},
archivePrefix = {arXiv},
       eprint = {1910.10154},
 primaryClass = {astro-ph.HE},
       adsurl = {https://ui.adsabs.harvard.edu/abs/2019arXiv191010154L}
}

@ARTICLE{guillochon2013,
       author = {{Guillochon}, James and {Ramirez-Ruiz}, Enrico},
        title = "{Hydrodynamical Simulations to Determine the Feeding Rate of Black Holes by the Tidal Disruption of Stars: The Importance of the Impact Parameter and Stellar Structure}",
      journal = {\apj},
         year = 2013,
        month = apr,
       volume = {767},
       number = {1},
          eid = {25},
        pages = {25},
          doi = {10.1088/0004-637X/767/1/25},
archivePrefix = {arXiv},
       eprint = {1206.2350},
 primaryClass = {astro-ph.HE},
       adsurl = {https://ui.adsabs.harvard.edu/abs/2013ApJ...767...25G}
}

@ARTICLE{paxton2011,
       author = {{Paxton}, Bill and {Bildsten}, Lars and {Dotter}, Aaron and {Herwig}, Falk and {Lesaffre}, Pierre and {Timmes}, Frank},
        title = "{Modules for Experiments in Stellar Astrophysics (MESA)}",
      journal = {\apjs},
         year = 2011,
        month = jan,
       volume = {192},
       number = {1},
          eid = {3},
        pages = {3},
          doi = {10.1088/0067-0049/192/1/3},
archivePrefix = {arXiv},
       eprint = {1009.1622},
 primaryClass = {astro-ph.SR},
       adsurl = {https://ui.adsabs.harvard.edu/abs/2011ApJS..192....3P}
}

@ARTICLE{bonnerot2021,
       author = {{Bonnerot}, Cl{\'e}ment and {Lu}, Wenbin and {Hopkins}, Philip F.},
        title = "{First light from tidal disruption events}",
      journal = {\mnras},
         year = 2021,
        month = jul,
       volume = {504},
       number = {4},
        pages = {4885-4905},
          doi = {10.1093/mnras/stab398},
archivePrefix = {arXiv},
       eprint = {2012.12271},
 primaryClass = {astro-ph.HE},
       adsurl = {https://ui.adsabs.harvard.edu/abs/2021MNRAS.504.4885B}
}
\begin{appendix}
    \section{Recovery via enthalpy conserving entropy}
    \label{app:recovery}
        Let us consider the known conserved variables in the computational frame $(\rho^*,S_i,e)$. 
        Then, to recover the primitive variables $(\rho,v^{\alpha},u)$, and $\Theta$, we need an equation of state for which we choose an ideal gas equation with an adiabatic index $\Gamma$, i.e. $P=(\Gamma-1)\rho u$. 
        Our baseline approach in SPHINCS is to write down an equation whose root is the new self-consistent pressure, and, once it is found by a numerical root finding, all other quantities can be recovered by a simple back-substitution. 
        This was first implemented for simple polytropes \citep{rosswog2021}, for piecewise polytropic equations of state with an ideal-gas type thermal contribution with an ideal gas-type thermal part \citep{rosswog2022}, for any type of cold equation of state together with physical thermal contributions from Fermi-liquid theory \cite{biswas2026} and for tabulated, nuclear equations of state \citep{shankar2026}.
        
        Motivated by \cite{springel2002}, \cite{liptai2019} formulated a relativistic entropy formulation to guarantee the positivity of the (specific) internal energy.
        Then, the pressure can be defined positive via
        \begin{equation}
            P=K\rho^\Gamma, \label{eq:polytrope}
        \end{equation}
        where $K$ is the (pseudo-)entropy variable. 
        Following \cite{tejeda2012} and \cite{liptai2019}, the recovery via enthalpy using the conservative variables $(\rho^*,S_i,K)$ implies solving the following implicit equation for the enthalpy
        \begin{equation}
            f(\mathcal{E})=1+\frac{\Gamma}{\Gamma-1}\frac{P(\mathcal{E})}{\rho(\mathcal{E})}-\mathcal{E}.
            \label{eq:f_enth}
        \end{equation}
        In order to find the expressions for $P=P(\mathcal{E})$ and $\rho(\mathcal{E})$  as functions strictly depending only on the conservative variables we use the following
        \begin{eqnarray}
            \rho(\mathcal{E})
            &=&
            \frac{\rho^*}{\sqrt{-g}\Theta(\mathcal{E})},
            \label{eq:rho_enth}
            \\
            P(\mathcal{E})
            &=&
            K\left(\frac{\rho^*}{\sqrt{-g}\Theta(\mathcal{E})}\right)^{\Gamma},
            \label{eq:P_enth}
            \\
            \Theta(\mathcal{E})
            &=&
            \frac{1}{\mathcal{E}}\sqrt{(g^{0i}S_i)^2-g^{00}(\mathcal{E}^2+S_iS^i)}, \label{eq:theta_enth}
            \\
            u(\mathcal{E})
            &=&
            \frac{P(\mathcal{E})}{(\Gamma-1)\rho(\mathcal{E})}.
            \label{eq:u_enth}
        \end{eqnarray}
        Here, equations~\ref{eq:rho_enth},~\ref{eq:P_enth}, and~\ref{eq:u_enth} are trivially derived from equations~\ref{eq:rhostar},~\ref{eq:polytrope}, and the ideal gas equation of state, respectively. 
        Here, it is relevant to clarify that we use a different convention for the definition of the generalised Lorentz factor. 
        As a result, equation~\ref{eq:theta_enth} differs from the expression for such a quantity seen in \cite{tejeda2012} and \cite{liptai2019}. 
        For details on its derivation, we defer the reader to Appendix~\ref{app:enthalpy}.

        In order to solve equation~\ref{eq:f_enth} using the Newton-Raphson method it is necessary to calculate its derivative with respect to the enthalpy, i.e.
        \begin{equation}
            \frac{df}{d\mathcal{E}}=\frac{\Gamma}{\Gamma-1}\frac{d}{d\mathcal{E}}\left(\frac{P}{\rho}\right)-1,
            \label{eq:dfdE} 
        \end{equation}
        %
\begin{comment}
        Applying the derivative of the product rule
         \begin{eqnarray}
            \frac{d}{d\mathcal{E}}\left(\frac{P}{\rho}\right)
            &=&
            \frac{1}{\rho}\frac{dP}{d\mathcal{E}}-\frac{P}{\rho^2}\frac{d\rho}{d\mathcal{E}},
            \\
            &=&
            -\frac{\Gamma}{\Theta}\frac{P}{\rho}\frac{d\Theta}{d\mathcal{E}}+\frac{P}{\rho\Theta}\frac{d\Theta}{d\mathcal{E}},
            \\
            &=&
            -\frac{\Gamma-1}{\Theta}\frac{P}{\rho}\frac{d\Theta}{d\mathcal{E}},
            \\
            &=&
            -\frac{\Gamma-1}{\Theta}\frac{P}{\rho}\frac{d}{d\mathcal{E}}\sqrt{\left(\frac{g^{0i}S_i}{\mathcal{E}}\right)^2-g^{00}\left(1+\frac{S_iS^i}{\mathcal{E}^2}\right)},
            \\
            &=&
            \frac{\Gamma-1}{\Theta^2}\frac{P}{\rho}\left[\frac{g^{00}S^iS_i-\left(g^{0i}S_i\right)^2}{\mathcal{E}^3}\right].
        \end{eqnarray}
        Replacing this quantity into equation~\ref{eq:dfdE} we obtain
        \begin{equation}
            \frac{df}{d\mathcal{E}}
            =
            \frac{\Gamma P}{\mathcal{E}^3\rho\Theta^2}\left[\left(g^{0i}S_i\right)^2-g^{00}S^iS_i\right]-1.
        \end{equation}
\end{comment}
        %
        which can be expressed as
        \begin{equation}
            \frac{df}{d\mathcal{E}}
            =
            \frac{\Gamma P}{\mathcal{E}^3\rho\Theta^2}\left[\left(g^{0i}S_i\right)^2-g^{00}S^iS_i\right]-1.
        \end{equation}
        Then, it is possible to apply the iterative Newton-Raphson method in order to calculate a solution of the enthalpy
        \begin{equation}
            \mathcal{E}^{n+1}=\mathcal{E}^n-f(\mathcal{E}^n)\left[\frac{df}{d\mathcal{E}}\left(\mathcal{E}^n\right)\right]^{-1},
        \end{equation}
        where the function $f(\mathcal{E})$ and its derivative are given in equations~\ref{eq:f_enth} and~\ref{eq:dfdE}, respectively. 
        At each iteration, the values of the functions $\rho(\mathcal{E})$, $P(\mathcal{E})$, and $\Theta(\mathcal{E})$ must be updated using equations~\ref{eq:rho_enth},~\ref{eq:P_enth}, and~\ref{eq:theta_enth}. 
        The iterative process continues until the relative residual falls below a preset tolerance that in our case is $\epsilon=10^{-13}$ as follows
        \begin{equation}
            \frac{\mathcal{E}^{n+1}-\mathcal{E}^n}{\mathcal{E}^{n+1}}<\epsilon.
        \end{equation}
        Once convergence is achieved, the primitive quantities $\rho$ and $u$, as well as $\Theta$ can be calculated using equations~\ref{eq:rho_enth},~\ref{eq:theta_enth}, and~\ref{eq:u_enth}, respectively. 
        The final step involves computing the velocity coordinates $v^{\alpha}$ from the conserved quantities, and the converged enthalpy value. 
        This can be done trivially as follows
        \begin{eqnarray}
            v_0
            &=&
            \frac{1}{g^{00}}\left(1 - \frac{g^{0i}S_i}{\Theta\mathcal{E}}\right),
            \\
            v_i
            &=&
            \frac{S_i}{\Theta\mathcal{E}}.
        \end{eqnarray}
        Then, it is just necessary to apply the contraviant tensor, so that $v^{\mu}=g^{\mu\nu}v_{\nu}$.
    \section{Derivation of the generalised Lorentz factor as a function of enthalpy}
    \label{app:enthalpy}
        Let us start from the definition of the generalised Lorentz factor
        \begin{equation}
            \Theta=\frac{1}{\sqrt{-g_{\mu\nu}v^{\mu}v^{\nu}}}.
            \label{eq:theta}
        \end{equation}
        As trivially $g_{\mu\nu}v^{\mu}v^{\nu}=g^{\mu\nu}v_{\mu}v_{\nu}$ we expand it as follows
        \begin{equation}
            g^{\mu\nu}v_{\mu}v_{\nu}=g^{00}+2g^{0i}v_0v_i+g^{ij}v_iv_j.
            \label{eq:g_equal}
        \end{equation}
        Additionally since $v^0=g^{0\mu}v_{\mu}=1$ (see equation~\ref{eq:coord_vel}), it is possible to write the following expression
        \begin{equation}
            v_0=\frac{1}{g^{00}}\left(1-g^{0i}v_i\right).
        \end{equation}
        Then,
        \begin{equation}
            g^{00}v_0^2=\frac{1}{g^{00}}\left[1-2g^{0i}v_i+\left(g^{0i}v_i\right)^2\right].
        \end{equation}
        Replacing these two expressions into equation~\ref{eq:g_equal}, we obtain
        \begin{equation}
            g^{\mu\nu}v_{\mu}v_{\nu}=\frac{1}{g^{00}}-\frac{\left(g^{0i}v_i\right)^2}{g^{00}}+g^{ij}v_iv_j.
        \end{equation}
        Let us apply the definition of the generalised Lorentz factor (see equation~\ref{eq:theta}),
        \begin{equation}
            \frac{1}{\Theta^2}=\left[\frac{\left(g^{0i}v_i\right)^2}{g^{00}}-\frac{1}{g^{00}}-g^{ij}v_iv_j\right],
        \end{equation}
        and multiply by $\mathcal{E}^2\Theta^2$ to solve for $\Theta^2$ in the following way
        \begin{equation}
            \Theta^2=\frac{1}{\mathcal{E}}\left[\left(\mathcal{E}\Theta g^{0i}v_i\right)^2-g^{00}\left(\mathcal{E}^2+\mathcal{E}^2\Theta^2g^{ij}v_iv_j\right)\right].
        \end{equation}
        Finally, inserting the canonical momentum $S_i=\Theta\mathcal{E}v_i$, we find
        \begin{equation}
            \Theta^2=\frac{1}{\mathcal{E}^2}\left[\left(g^{0i}S_i\right)^2-g^{00}\left(\mathcal{E}^2+g^{ij}S_iS_j\right)\right],
        \end{equation}
        which, after applying square root, corresponds to the generalised Lorentz factor $\Theta$ as a function of the metric $g$, the conservative quantities $S_i$, and enthalpy $\mathcal{E}$ as shown in equation~\ref{eq:theta_enth}.
\section{Kerr metric in CKS coordinates}
\label{app:metric}
    Following \cite{kerr1963}, the components of the Kerr-Schild covariant metric tensor in Cartesian-like coordinates $(t,x,y,z)$ considering a central object of mass $M$ and spin $a$ is given by
    \begin{eqnarray}
        g_{tt}(t,x,y,z)&=&-1+\frac{2Mr}{\rho^2_\text{met}},
        \\
        g_{xx}(t,x,y,z)&=&1+\frac{2Mr}{\rho^2_\text{met}}\left(\frac{rx+ay}{r^2+a^2}\right)^2,
        \\
        g_{yy}(t,x,y,z)&=&1+\frac{2Mr}{\rho^2_\text{met}}\left(\frac{ry-ax}{r^2+a^2}\right)^2,
        \\
        g_{zz}(t,x,y,z)&=&1+\frac{2Mz^2}{r\rho^2_\text{met}},
        \\
        g_{tx}(t,x,y,z)&=&\frac{2Mr}{\rho^2_\text{met}}\left(\frac{rx+ay}{r^2+a^2}\right),
        \\
        g_{ty}(t,x,y,z)&=&\frac{2Mr}{\rho^2_\text{met}}\left(\frac{ry-ax}{r^2+a^2}\right),
        \\
        g_{tz}(t,x,y,z)&=&\frac{2Mz}{\rho^2_\text{met}},
        \\
        g_{xz}(t,x,y,z)&=&\frac{2Mz}{\rho^2_\text{met}}\left(\frac{rx+ay}{r^2+a^2}\right),
        \\
        g_{yz}(t,x,y,z)&=&\frac{2Mz}{\rho^2_\text{met}}\left(\frac{ry-ax}{r^2+a^2}\right),
        \\
        g_{xy}(t,x,y,z)&=&\frac{2Mr}{\rho^2_\text{met}}\frac{(rx+ay)(ry-ax)}{(r^2+a^2)^2},
    \end{eqnarray}
    where
    \begin{equation}
        \rho_\text{met}=r^2+\frac{a^2z^2}{r^2}=\sqrt{(x^2+y^2+z^2-a^2)^2+4a^2z^2}.
        \label{eq:rho_metric_app}
    \end{equation}
\section{Derivation of the CKS Kerr metric derivatives}
\label{app:dmetric}
    In order to calculate the right-hand side of the equation~(\ref{eq:momentum}) 
    we need the derivatives of the covariant metric tensor $g_{\mu\nu}$ with respect to the spatial coordinates $x^i$.   
    In the Phantom code \citep{liptai2019}, these derivatives are calculated with a 2nd-order accurate Finite Difference approach but they can, of course, also be calculated analytically which is what we do here.
    First, let us consider the set of CKS coordinates $(t,x,y,z)$, and the definitions of $r$ and $\rho_\text{met}$ from equation~\ref{eq:rho_metric_app}. 
    Then, 
    \begin{eqnarray}
        \frac{\partial(r^2)}{\partial x}
        &=&
        x+\frac{x(x^2+y^2+z^2-a^2)}{\sqrt{(x^2+y^2+z^2-a^2)^2+4a^2z^2}}
        =
        \frac{2xr^2}{\rho^2_\text{met}},
        \\
        \frac{\partial(r^2)}{\partial y}
        &=&
        y+\frac{y(x^2+y^2+z^2-a^2)}{\sqrt{(x^2+y^2+z^2-a^2)^2+4a^2z^2}}
        =
        \frac{2yr^2}{\rho^2_\text{met}},
        \\
        \frac{\partial(r^2)}{\partial z}
        &=&
        z+\frac{z(x^2+y^2+z^2-a^2)+2a^2z}{\sqrt{(x^2+y^2+z^2-a^2)^2+4a^2z^2}}
        =
        \frac{2z(r^2+a^2)}{\rho^2_\text{met}},
        \nonumber
        \\
        &&
    \end{eqnarray}
    and trivially,
    \begin{eqnarray}
        \frac{\partial r}{\partial x}
        &=&
        \frac{1}{2r}\frac{\partial (r^2)}{\partial x}
        =
        \frac{xr}{\rho^2_\text{met}},
        \\
        \frac{\partial r}{\partial y}
        &=&
        \frac{1}{2r}\frac{\partial (r^2)}{\partial y}
        =
        \frac{yr}{\rho^2_\text{met}},
        \\
        \frac{\partial r}{\partial z}
        &=&
        \frac{1}{2r}\frac{\partial (r^2)}{\partial z}
        =
        \frac{z(r^2+a^2)}{r\rho^2_\text{met}}.
    \end{eqnarray}
    Then, for the $\rho^2_\text{met}$-derivatives,
    \begin{eqnarray}
        \frac{\partial(\rho^2_\text{met})}{\partial x}
        &=&
        \left(1-\frac{a^2z^2}{r^4}\right)\frac{\partial(r^2)}{\partial x}
        \\
        &=&
        \frac{2xr^2}{\rho^2_\text{met}}\left(1-\frac{a^2z^2}{r^4}\right)
        \\
        \frac{\partial(\rho^2_\text{met})}{\partial y}
        &=&
        \left(1-\frac{a^2z^2}{r^4}\right)\frac{\partial(r^2)}{\partial y}
        \\
        &=&
        \frac{2yr^2}{\rho^2_\text{met}}\left(1-\frac{a^2z^2}{r^4}\right)
        \\
        \frac{\partial(\rho^2_\text{met})}{\partial z}
        &=&
        \left(1-\frac{a^2z^2}{r^4}\right)\frac{\partial(r^2)}{\partial z}+\frac{2a^2z}{r^2}
        \\
        &=&
        \frac{2z(r^2+a^2)}{\rho^2_\text{met}}\left(1-\frac{a^2z^2}{r^4}\right)+\frac{2a^2z}{r^2}.
    \end{eqnarray}
    Additionally, we calculate the following derivatives
    \begin{eqnarray}
%        \frac{\partial}{\partial x}\left(\frac{rx+ay}{r^2+a^2}\right)
%        &=&
%        \frac{1}{(r^2+a^2)^2}\left[(r^2+a^2)\frac{\partial}{\partial x}(rx+ay)-(rx+ay)\frac{\partial}{\partial x}(r^2+a^2)\right]
%        \\
%        &=&
%        \frac{1}{(r^2+a^2)^2}\left[(r^2+a^2)\left(x\frac{\partial r}{\partial x}+r\right)-(rx+ay)\frac{\partial (r^2)}{\partial x}\right]
%        \\
        \frac{\partial}{\partial x}\left(\frac{rx+ay}{r^2+a^2}\right)
        &=&
        \frac{1}{r^2+a^2}\left[r+\frac{\partial r}{\partial x}\left(x-2r\frac{rx+ay}{r^2+a^2}\right)\right]
        \\
        \Rightarrow f_x
        &=&
        \left[r+r_x\left(x-2rf\right)\right]/(r^2+a^2)
        \\
        \frac{\partial}{\partial y}\left(\frac{rx+ay}{r^2+a^2}\right)
        &=&
        \frac{1}{r^2+a^2}\left[a+\frac{\partial r}{\partial y}\left(x-2r\frac{rx+ay}{r^2+a^2}\right)\right]
        \\
        \Rightarrow f_y
        &=&
        \left[a+r_y\left(x-2rf\right)\right]/(r^2+a^2)
        \\
        \frac{\partial}{\partial z}\left(\frac{rx+ay}{r^2+a^2}\right)
        &=&
        \frac{1}{r^2+a^2}\left[\frac{\partial r}{\partial z}\left(x-2r\frac{rx+ay}{r^2+a^2}\right)\right]
        \\
        \Rightarrow f_z
        &=&
        \left[r_z\left(x-2rf\right)\right]/(r^2+a^2)
    \end{eqnarray}
    \begin{eqnarray}
        \frac{\partial}{\partial x}\left(\frac{ry-ax}{r^2+a^2}\right)
        &=&
        \frac{1}{r^2+a^2}\left[-a+\frac{\partial r}{\partial x}\left(y-2r\frac{ry-ax}{r^2+a^2}\right)\right]
        \\
        \Rightarrow h_x
        &=&
        \left[-a+r_y\left(y-2rh\right)\right]/(r^2+a^2)
        \\
        \frac{\partial}{\partial y}\left(\frac{ry-ax}{r^2+a^2}\right)
        &=&
        \frac{1}{r^2+a^2}\left[r+\frac{\partial r}{\partial y}\left(y-2r\frac{ry-ax}{r^2+a^2}\right)\right]
        \\
        \Rightarrow h_y
        &=&
        \left[r+r_y\left(y-2rh\right)\right]/(r^2+a^2)
        \\
        \frac{\partial}{\partial z}\left(\frac{ry-ax}{r^2+a^2}\right)
        &=&
        \frac{1}{r^2+a^2}\left[\frac{\partial r}{\partial z}\left(y-2r\frac{ry-ax}{r^2+a^2}\right)\right]
        \\
        \Rightarrow h_z
        &=&
        \left[r_z\left(y-2rh\right)\right]/(r^2+a^2)
    \end{eqnarray}
    Now, let us define $C_M=2M/\rho^4_\text{met}$
    \begin{eqnarray}
        \frac{\partial}{\partial x}\left(\frac{2Mr}{\rho^2_\text{met}}\right)
        &=&
        \frac{2M}{\rho^4}\left[\rho^2_\text{met}\frac{\partial r}{\partial x}-r\frac{\partial(\rho^2_\text{met})}{\partial x}\right]
        \\
        \Rightarrow
        (g_{tt}+1)_x
        &=&
        C_M\left[\rho^2r_x-r(\rho^2_\text{met})_x\right]
        \\
        \frac{\partial}{\partial y}\left(\frac{2Mr}{\rho^2_\text{met}}\right)
        &=&
        \frac{2M}{\rho^4_\text{met}}\left[\rho^2\frac{\partial r}{\partial y}-r\frac{\partial(\rho^2_\text{met})}{\partial y}\right]
        \\
        \Rightarrow
        (g_{tt}+1)_y
        &=&
        C_M\left[\rho^2_\text{met}r_y-r(\rho^2_\text{met})_y\right]
        \\
        \frac{\partial}{\partial z}\left(\frac{2Mr}{\rho^2}\right)
        &=&
        \frac{2M}{\rho^4_\text{met}}\left[\rho^2_\text{met}\frac{\partial r}{\partial z}-r\frac{\partial(\rho^2_\text{met})}{\partial z}\right]
        \\
        \Rightarrow
        (g_{tt}+1)_z
        &=&
        C_M\left[\rho^2r_z-r(\rho^2_\text{met})_z\right]
    \end{eqnarray}
    \begin{eqnarray}
        \frac{\partial}{\partial x}\left(\frac{2Mz^2}{r\rho^2_\text{met}}\right)
        &=&
        -\frac{2M}{\rho^4}\frac{z^2}{r^2}\left(\rho^2\frac{\partial r}{\partial x}+r\frac{\partial(\rho^2_\text{met})}{\partial x}\right)
        \\
        \Rightarrow
        (g_{zz})_x
        &=&
        -C_M\frac{z^2}{r^2}\left[\rho^2_\text{met}r_x+r(\rho^2_\text{met})_x\right]
        \\
        \frac{\partial}{\partial y}\left(\frac{2Mz^2}{r\rho^2_\text{met}}\right)
        &=&
        -\frac{2M}{\rho^4_\text{met}}\frac{z^2}{r^2}\left(\rho^2\frac{\partial r}{\partial y}+r\frac{\partial(\rho^2_\text{met})}{\partial y}\right)
        \\
        \Rightarrow
        (g_{zz})_y
        &=&
        -C_M\frac{z^2}{r^2}\left[\rho^2r_y+r(\rho^2_\text{met})_y\right]
        \\
        \frac{\partial}{\partial z}\left(\frac{2Mz^2}{r\rho^2_\text{met}}\right)
        &=&
        \frac{2M}{\rho^4_\text{met}}\frac{1}{r^2}\left[2zr\rho^2-z^2\left(\rho^2_\text{met}\frac{\partial r}{\partial z}+r\frac{\partial(\rho^2_\text{met})}{\partial z}\right)\right]
        \nonumber
        \\
        &&
        \\
        \Rightarrow
        (g_{zz})_z
        &=&
        C_M\frac{1}{r^2}\left[2zr\rho^2_\text{met}-z^2\left(\rho^2_\text{met}r_z+r(\rho^2_\text{met})_z\right)\right]
        \nonumber
        \\
        &&
    \end{eqnarray}
    \begin{eqnarray}
        \frac{\partial}{\partial x}\left(\frac{2Mz}{\rho^2_\text{met}}\right)
        &=&
        -\frac{2M}{\rho^4_\text{met}}\left[z\frac{\partial(\rho^2_\text{met})}{\partial x}\right]
        \\
        \Rightarrow (g_{tz})_x
        &=&
        -C_M\left[z(\rho^2_\text{met})_x\right]
        \\
        \frac{\partial}{\partial y}\left(\frac{2Mz}{\rho^2_\text{met}}\right)
        &=&
        -\frac{2M}{\rho^4_\text{met}}\left[z\frac{\partial(\rho^2_\text{met})}{\partial y}\right]
        \\
        \Rightarrow (g_{tz})_y
        &=&
        -C_M\left[z(\rho^2_\text{met})_y\right]
        \\
        \frac{\partial}{\partial z}\left(\frac{2Mz}{\rho^2_\text{met}}\right)
        &=&
        \frac{2M}{\rho^4}\left[\rho^2_\text{met}-z\frac{\partial(\rho^2_\text{met})}{\partial z}\right]
        \\
        \Rightarrow (g_{tz})_z
        &=&
        C_M\left[\rho^2-z(\rho^2_\text{met})_z\right].
    \end{eqnarray}
    
    Then, the derivatives of the metric with respect to $x$ are
    \begin{eqnarray}
        \frac{\partial g_{tt}}{\partial x}
        &=&
        C_M\left[\rho^2r_x-r\left(\rho^2_\text{met}\right)_x\right],
        \\
        \frac{\partial g_{tx}}{\partial x}
        &=&
        (g_{tt}+1)_xf+(g_{tt}+1)f_x,
        \\
        \frac{\partial g_{ty}}{\partial x}
        &=&
        (g_{tt}+1)_xh+(g_{tt}+1)h_x,
        \\
        \frac{\partial g_{tz}}{\partial x}
        &=&
        -C_M\left[z(\rho^2_\text{met})_x\right],
        \\
        \frac{\partial g_{xx}}{\partial x}
        &=&
        (g_{tt}+1)_xf^2+(g_{tt}+1)(f^2)_x,
        \\
        \frac{\partial g_{xy}}{\partial x}
        &=&
        (g_{tt}+1)_xfh+(g_{tt}+1)(f_xh+fh_x),
        \\
        \frac{\partial g_{xz}}{\partial x}
        &=&
        (g_{tz})_xf+g_{tz}f_x,
        \\
        \frac{\partial g_{yy}}{\partial x}
        &=&
        (g_{tt}+1)_xh^2+(g_{tt}+1)(h^2)_x,
        \\
        \frac{\partial g_{yz}}{\partial x}
        &=&
        (g_{tz})_xh+g_{tz}h_x,
        \\
        \frac{\partial g_{zz}}{\partial x}
        &=&
        -C_M\left(z^2/r^2\right)\left[\rho^2_\text{met}r_x+r(\rho^2_\text{met})_x\right],
    \end{eqnarray}
    with respect to $y$
    \begin{eqnarray}
        \frac{\partial g_{tt}}{\partial y}
        &=&
        C_M\left[\rho^2_\text{met}r_y-r\left(\rho^2_\text{met}\right)_y\right],
        \\
        \frac{\partial g_{tx}}{\partial y}
        &=&
        (g_{tt}+1)_yf+(g_{tt}+1)f_y,
        \\
        \frac{\partial g_{ty}}{\partial y}
        &=&
        (g_{tt}+1)_yh+(g_{tt}+1)h_y,
        \\
        \frac{\partial g_{tz}}{\partial y}
        &=&
        -C_M\left[z(\rho^2)_y\right],
        \\
        \frac{\partial g_{xx}}{\partial x}
        &=&
        (g_{tt}+1)_yf^2+(g_{tt}+1)(f^2)_y,
        \\
        \frac{\partial g_{xy}}{\partial y}
        &=&
        (g_{tt}+1)_yfh+(g_{tt}+1)(f_yh+fh_y),
        \\
        \frac{\partial g_{xz}}{\partial y}
        &=&
        (g_{tz})_yf+g_{tz}f_y,
        \\
        \frac{\partial g_{yy}}{\partial y}
        &=&
        (g_{tt}+1)_yh^2+(g_{tt}+1)(h^2)_y,
        \\ 
        \frac{\partial g_{yz}}{\partial y}
        &=&
        (g_{tz})_yh+g_{tz}h_y,
        \\
        \frac{\partial g_{zz}}{\partial y}
        &=&
        -C_M\left(z^2/r^2\right)\left[\rho^2_\text{met}r_y+r(\rho^2_\text{met})_y\right],
    \end{eqnarray}
    and with respect to $z$ are
    \begin{eqnarray}
        \frac{\partial g_{tt}}{\partial z}
        &=&
        C_M\left[\rho^2_\text{met}r_z-r\left(\rho^2_\text{met}\right)_z\right],
        \\
        \frac{\partial g_{tx}}{\partial z}
        &=&
        (g_{tt}+1)_zf+(g_{tt}+1)f_z,
        \\
        \frac{\partial g_{ty}}{\partial z}
        &=&
        (g_{tt}+1)_zh+(g_{tt}+1)h_z,
        \\
        \frac{\partial g_{tz}}{\partial z}
        &=&
        C_M\left[\rho^2-z(\rho^2)_z\right],
        \\
        \frac{\partial g_{xx}}{\partial z}
        &=&
        (g_{tt}+1)_zf^2+(g_{tt}+1)(f^2)_z,
        \\
        \frac{\partial g_{xy}}{\partial z}
        &=&
        (g_{tt}+1)_zfh+(g_{tt}+1)(f_zh+fh_z),
        \\
        \frac{\partial g_{xz}}{\partial z}
        &=&
        (g_{tz})_zf+g_{tz}f_z,
        \\
        \frac{\partial g_{yy}}{\partial z}
        &=&
        (g_{tt}+1)_zh^2+(g_{tt}+1)(h^2)_z,
        \\
        \frac{\partial g_{yz}}{\partial z}
        &=&
        (g_{tz})_zh+g_{tz}h_z,
        \\
        \frac{\partial g_{zz}}{\partial z}
        &=&
        C_M\left(1/r^2\right)\left[2zr\rho^2-z^2\left(\rho^2_\text{met}r_z+r(\rho^2_\text{met})_z\right)\right].
    \end{eqnarray}
\section{Validation tests}
\label{app:tests}
    Here we present a series of tests to demonstrate the correct working of our implementation into SPHINCS. 
    The presented tests are inspired by work that have validated particle dynamics in time-independent metrics, including Kerr metric \citep[e.g.][]{liptai2019,lupi2023}. 
    Unless stated otherwise, the tests were performed using a fixed timestep of $\Delta t=0.01~\text{M}$.
        \begin{figure}
            \centering
            \includegraphics[width=0.95\linewidth]{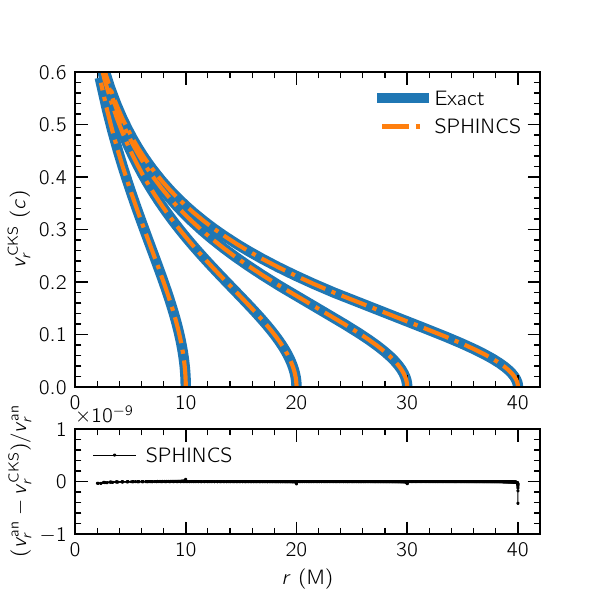}
            \caption{
            Velocity as a function of distance for four simulations with initial distance $r_0=10$,~$20$,~$30$,~$40$~M. 
            The solid blue and dot-dashed orange lines show the analytical and simulation calculations, respectively. 
            The bottom panel shows the relative residual of comparing the analytical and simulated results. 
            The calculations agreement is overall $<$10$^{-9}$.
            }
            \label{fig:ff}
        \end{figure}
        \begin{figure}
            \centering
            \includegraphics[width=0.8\linewidth]{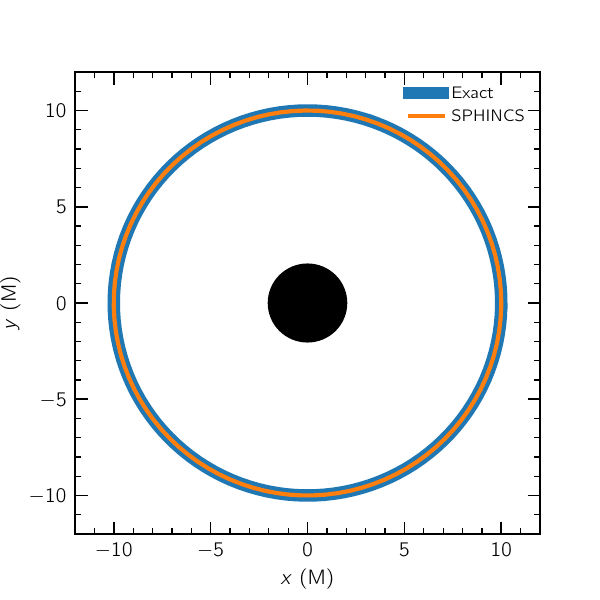}
            \includegraphics[width=0.8\linewidth]{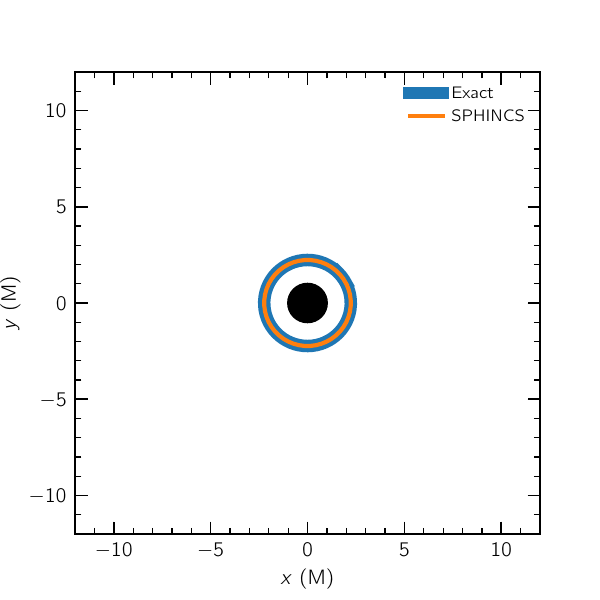}
            \caption{
            Stable circular orbit tests for a particle in Kerr metric with $M=1$ combined with $a=0$ (top panel) and $a=0.998$ (bottom panel). 
            The exact geodesics and the simulation results are shown as thick blue and thin orange lines, respectively.
            }
            \label{fig:circular}
        \end{figure}
        \begin{figure*}
            \centering
            \includegraphics[width=0.85\linewidth]{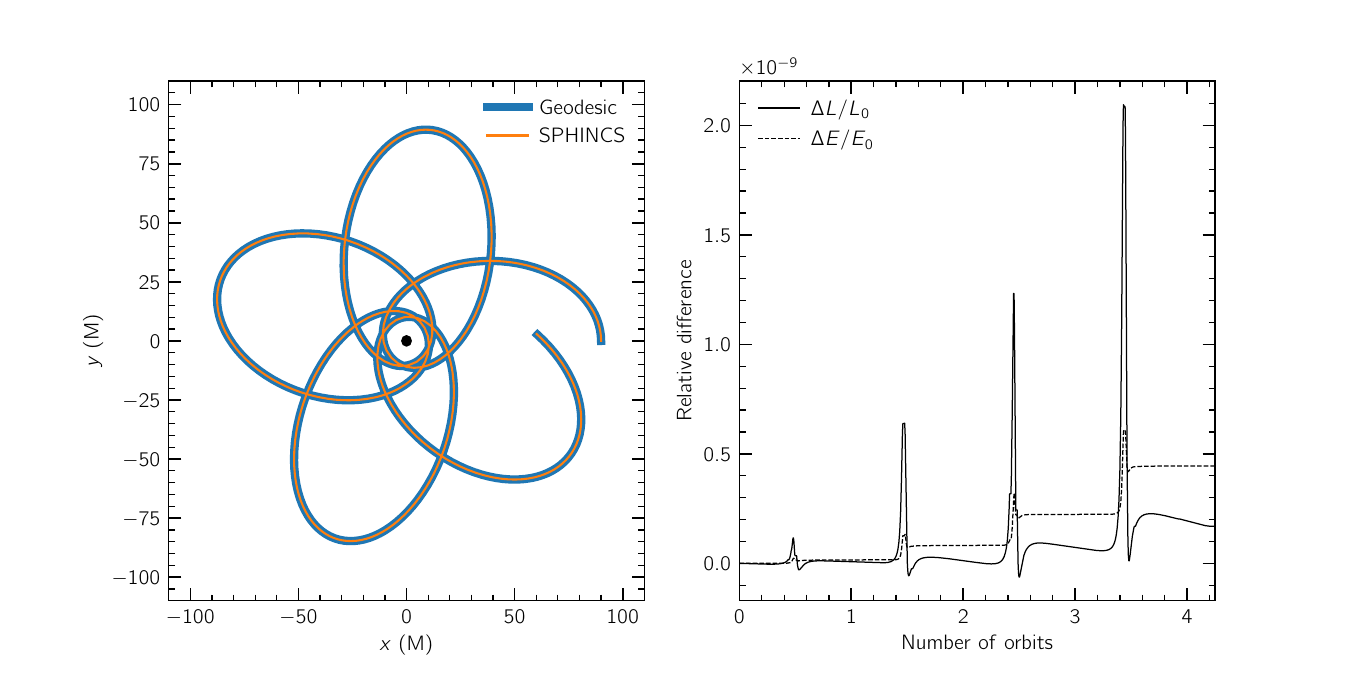}
            \caption{
            Precession test in Kerr metric with $M=1$ and $a=0$. 
            The left-hand side panel shows the geodesic and the simulated result in the $xy$ plane as thick blue and thin orange lines, respectively. 
            The right-hand side panel show the relative changes in specific angular momentum (solid line) and specific energy (dashed line) as functions of the number of orbits. 
            Notice that at most the relative differences are of the order of $10^{-9}$.
            }
            \label{fig:precession}
        \end{figure*}
        \begin{figure*}
            \centering
            \includegraphics[width=0.85\linewidth]{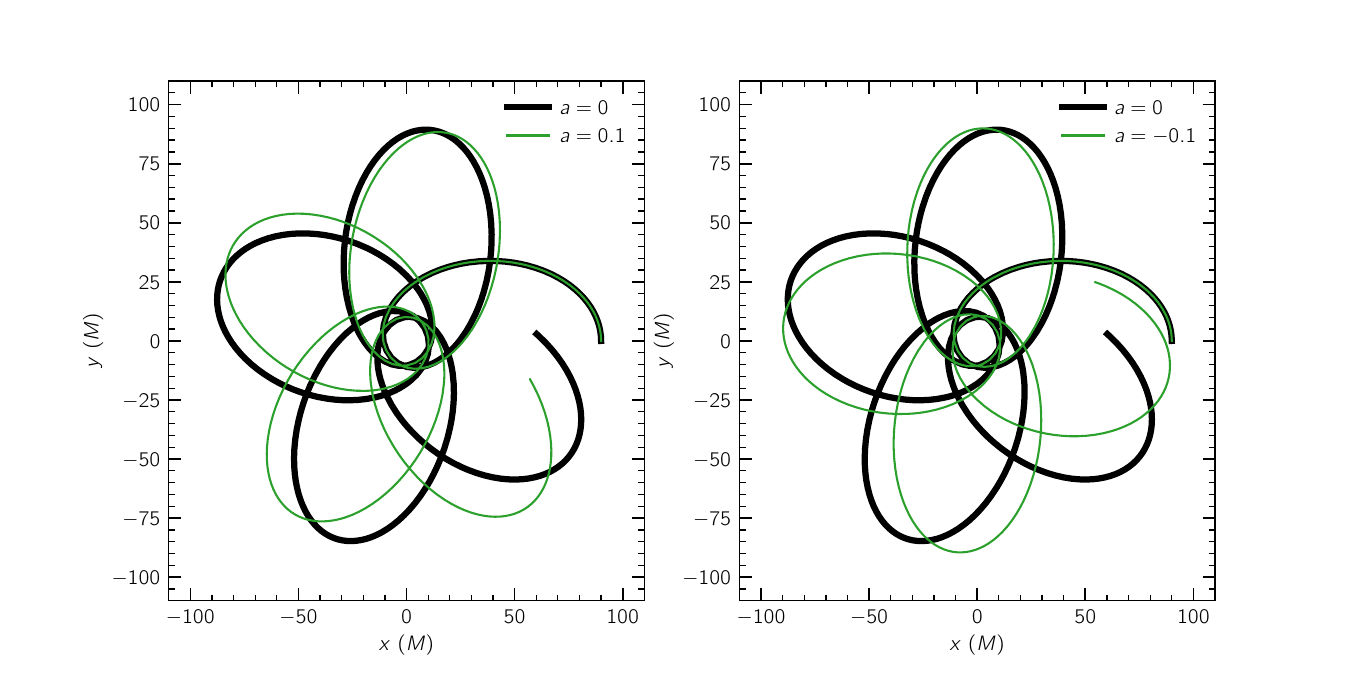}
            \caption{
            Precession tests in Kerr metric with $M=1$ with $a=0.1$ (left-hand side) and $a=-0.1$ (right-hand side). 
            The simulated results with $a=0$ and $a=\pm0.1$ are shown as thick black and thin green lines, respectively.}
            \label{fig:shifts}
        \end{figure*}
        \begin{figure*}
            \centering
            \includegraphics[width=\linewidth]{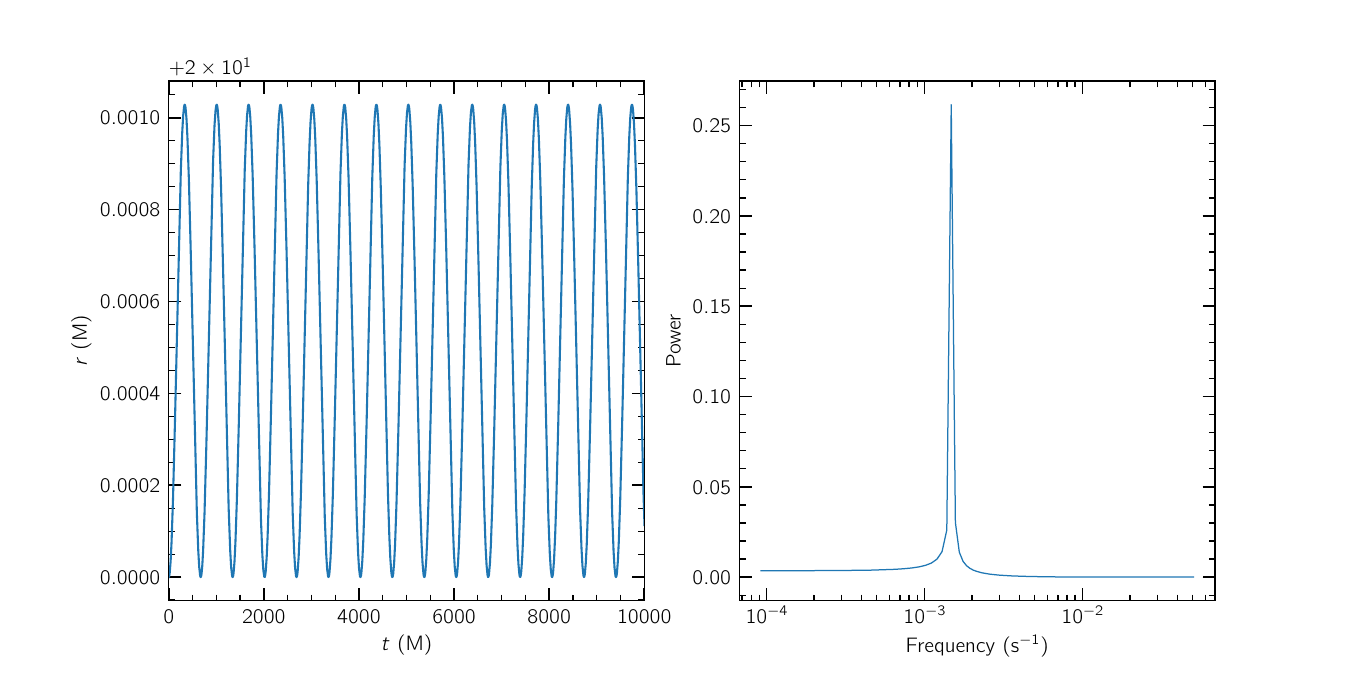}
            \caption{
            Analysis of the epicyclic motion test in Kerr metric with $M=1$ and $a=0$ for a test particle initially at $r=20$~M. 
            The left-hand side panel shows the radial distance as a function of time, displaying the epicyclic motion. 
            The right-hand side panel contains the Fast-Fourier transform of the periodic signal that was used to obtain the frequency of the epicyclic motion.
            }
            \label{fig:fft}
        \end{figure*}
        \begin{figure}
            \centering
            \includegraphics[width=0.9\linewidth]{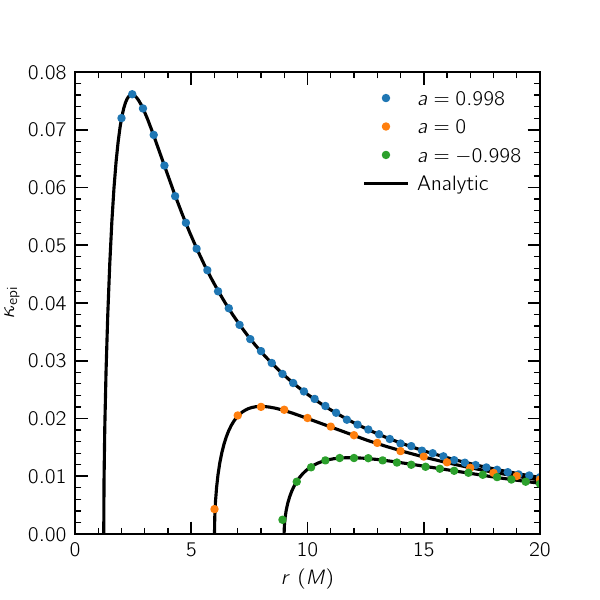}
            \caption{
            Epicyclic motion test in Kerr metric with $M=1$. 
            The epicyclic frequencies $\kappa_\text{epi}$ are shown as functions of the initial radial distance as a solid black line.
            Blue, orange, and green dots represent the result of the numerical calculations for the cases with $a=-0.998$,~$0$,~$0.998$, respectively.
            The solid black lines correspond to the analytic expression shown in equation~\ref{eq:epi}.
            }
            \label{fig:epi}
        \end{figure}
    \subsection{Radial geodesics in Kerr metric}
        We first place a test particle at a distance $r_0$ from a central object of mass $M$ and spin $a=0$. 
        The object is initially at rest, and starts moving due to the curvature of the metric. 
        Then, the radial velocity of such a particle will be determined by its position $r$. 
        The solution in Schwarzschild coordinates is given as
        \begin{equation}
            v^r(r)=\frac{1-\frac{2M}{r}}{\sqrt{1-\frac{2M}{r_0}}}\sqrt{2M\left(\frac{1}{r}-\frac{1}{r_0}\right)}.
        \end{equation}
        We tested our SPHINCS implementation in reproducing
        this test case by placing test particles at initial distances $r_0=10$,~$20$,~$30$,~$40$~M at rest. 
        The evolution was calculated until the particle reached a distance of $r=2$~M. 
        The results of these calculations are shown in Fig.~\ref{fig:ff} displaying the radial velocity as a function of the distance. 
        Additionally, we included the analytical solution in CKS coordinates in order to compare the agreement between them (see Appendix~\ref{app:coordinates}). 
        Notice that the agreement is about $\sim$10$^{-10}$ which results in a successful test.
    \subsection{Stable circular orbits in Kerr metric}
        In order to test the ability of SPHINCS to simulate stable circular orbits we simulated two cases: a particle at $r=10$~M orbiting central object of $M=1$ and $a=0$, and another particle at $r=2$~M moving around an object with $M=1$ and $a=0.998$. 
        The angular velocity of such orbits in Kerr metric \citep{abramowicz1978} given by
        \begin{equation}
            \Omega=\frac{M^{1/2}}{r^{-3/2}+aM^{1/2}}.
        \end{equation}
        Our results are shown in Figure~\ref{fig:circular}, where the cases with $a=0$ and $a=0.998$ are displayed in the top and bottom panels, respectively. 
        Within 20 orbits, the energy and angular momentum were conserved to $\sim$10$^{-11}$ for the non-spinning case and $\sim$10$^{-7}$ for the rapidly spinning case.
    \subsection{Orbital shift in Kerr metric}
        In order to test the ability of the code to capture geodesics and conservation of energy (and momentum) we set up a particle at $x=90$~M, $y=0$, $z=0$, and $v^x=0$, $v^y=0.0521157$, $v^z=0$ moving due to the presence of a central object with $M=1$ and $a=0$. 
        The motion along the orbital plane $xy$ is shown on the left-hand side panel of Figure~\ref{fig:precession}. 
        The solid orange line shows the result of the simulation while the solid thick blue line represents the expected geodesic. 
        On the right-hand side panel, we show the relative difference of the magnitude of the angular momentum and energy compared to the initial values of the particle as a function of the number of orbits. 
        Note that the quantities are conserved to $\sim$10$^{-9}$.

        For testing the effect of spin on geodesics we repeated the previous test but setting $a=\pm0.1$. 
        Figure~\ref{fig:shifts} show the results where the cases with $a=+0.1$ and $a=-0.1$ are displayed on the left- and right-hand side panels. 
        The panels show the geodesics computed from the simulation as solid green lines and the non-spinning case is shown as reference as a solid black line. 
        These results show the expected precession due to the spin of the central object.
    \subsection{Epicyclic motion}
        If a small, radial perturbation is introduced into a circular orbital motion in Kerr metric the result will be epicyclic motion, whose frequency can be described analytically \citep{kato1990,lubow2002}. 
        Then, the epicyclic frequency is given by 
        \begin{equation}
            \label{eq:epi}
            \kappa^2_\text{epi}=\Omega^2\left(1-\left[\frac{6M}{r}-\frac{8aM^{1/2}}{r^{3/2}}+\frac{3a^2}{r^2}\right]\right).
        \end{equation}
        Here we tested if SPHINCS can capture correctly this epicyclic motion. 
        We performed the evolution of a set of test particles at different separation from the central object with $M=1$ and the cases of $a=-0.998$, $0$, $+0.998$. 
        The particles were placed initially at $r_0\leq20$~M on a circular orbit. 
        A small perturbation was added setting the initial angular velocity to be 1.00001$\Omega$, and evolved up to $t=10000$~M.
        Figure~\ref{fig:fft} shows an example of the evolution for a particle initially at $r_0=20$~M. 
        The left-hand side panel contains the radial distance as a function of time, and the right-hand side panel displays the Fast-Fourier transform of the periodic motion observed on the left-hand side panel. 
        The results of the complete analysis of this procedure is shown in Figure~\ref{fig:epi} that shows the epicyclic frequency $\kappa_\text{epi}$ as a function of the initial radial distance. 
        The frequencies computed from the numerical simulation are represented as blue, orange, and green points for the cases with $a=-0.998,0,+0.998$, respectively. 
        As a reference, we show the analytical expression of the epicyclic frequency from equation~\ref{eq:epi} as a solid black line. 
        Here it is possible to see that the epicyclic motion is well described in the simulations as we find the expected epicyclic frequencies even when considering rapidly rotating central objects.
        \begin{table*}
            \centering
            \begin{threeparttable}
            \caption{
            Fitting parameters of the properties of the mass fallback rates.
            }
            \begin{tabular}{lccc}
                \hline
                \\
                Quantity
                &
                Range of $\beta$
                &
                Knots
                &
                Coefficients
                \\
                (1) & (2) & (3) & (4)
                \\
                \hline
                \\
                $A_{5/3,\text{K}}^{0}$
                &
                $[1,8]$
                &
                $[1,2,4,6,8]$
                &
                $[1.793, 1.339, 0.8122, 0.5978, 0.5181]$
                \\
                $A_{5/3,\text{K}}^{+}$
                &
                $[1,10]$
                &
                $[1,2,4,6,8,10]$
                &
                $[1.771, 1.381, 0.8856, 0.6906, 0.5986, 0.5421]$
                \\
                $A_{5/3,\text{K}}^{-}$
                &
                $[1,6]$
                &
                $[1,2,4,6]$
                &
                $[1.812, 1.288, 0.7530, 0.5563]$
                \\
                \\
                $B_{5/3}^0$
                &
                $[1,8]$
                &
                $[1,2,4,6,8]$
                &
                $[0.1451, 0.1592, 0.2105, 0.2105, 1.748]$
                \\
                $B_{5/3}^+$
                &
                $[1,10]$
                &
                $[1,2,4,6,8,10]$
                &
                $[0.1451, 0.1592, 0.2105, 0.2310, 0.1918, 0.1748]$
                \\
                $B_{5/3}^-$
                &
                $[1,6]$
                &
                $[1,2,4,6]$
                &
                $[0.1451, 0.1592, 0.2105, 0.1918]$
                \\
                \\
                $D_{5/3}^0$
                &
                $[1,8]$
                &
                $[1,2,4,6,8]$
                &
                $[-1.570, -1.684, -1.696, -1.632, -1.583]$
                \\
                $D_{5/3}^+$
                &
                $[1,10]$
                &
                $[1,2,4,6,8,10]$
                &
                $[-1.560, -1.685, -1.691, -1.682, 1.625, -1.584]$
                \\
                $D_{5/3}^-$
                &
                $[1,6]$
                &
                $[1,2,4,6]$
                &
                $[-1.581, -1.686, -1.712, -1.602]$
                \\
                \\
                $K_{5/3}^0$
                &
                $[1,8]$
                &
                $[1,2,4,6,8]$
                &
                $[0.02420,0.02578,0.03699,0.03734,0.03666]$
                \\
                $K_{5/3}^+$
                &
                $[1,10]$
                &
                $[1,2,4,6,8,10]$
                &
                $[0.02413, 0.02594,0.03840, 0.04781, 0.04675, 0.04367]$
                \\
                $K_\text{5/3}^-$
                &
                $[1,6]$
                &
                $[1,2,4,6]$
                &
                $[0.02439, 0.02529, 0.03253,0.03320]$
                \\
                \\
                \hline
                \hline
            \end{tabular}
            \label{tab:fits}
            \begin{tablenotes}
                \item
                \textit{Notes.} 
                Column~1: fitted quantity.
                Column~2: $\beta$ range. 
                Column~3: discrete knots of the fits.
                 Column~4: coefficients of the fits. 
            \end{tablenotes}
        \end{threeparttable}
        \end{table*}
    \section{Parabolic orbit in Kerr metric}
    \label{sec:parabolic}
        Let us consider an equatorial parabolic orbit, i.e. with specific energy $E=1$, with a pericentre distance $r_\text{p}$ around a central object of mass $M$ and spin $a$ in Kerr metric. 
        Then, the specific angular momentum component $\ell_z$ in Boyer–Lindquist coordinates is 
        \begin{equation}
            \ell_z=\frac{-2aM\pm\sqrt{2Mr_\text{p}\Delta_\text{p}}}{r_\text{p}-2M},
        \end{equation}
        where $\Delta_\text{p}=r_\text{p}^2-2Mr_\text{p}+a^2$, and the plus or minus signs indicate if the orbit is prograde or retrograde with respect to the central object spin, respectively. 
        Notice that if $a=0$, $\Delta_\text{p}=r_\text{p}^2-2Mr_\text{p}$ and the expression reduces to the Schwarzschild case as expected, i.e.
        \begin{equation}
            \ell_z=\pm\sqrt{\frac{2Mr_\text{p}^2}{r_\text{p}-2M}},
        \end{equation}
        where the signs ``$+$" and ``$-$" in both expressions determines if the orbital motion is counter clockwise or clockwise, respectively.
    \section{Mass fallback rate fitting parameters}
    \label{app:fitting}
        In Section~\ref{sec:fallback}, we have characterised the mass fallback rates of each model by calculating their peak fallback rate, the time of the peak, the long-term time decay power law $n_\infty$, and the rise-to-peak timescale $\tau_\text{rise}$. 
        We have fitted B-spline functions to these quantities as functions of the impact factor $\beta$ as shown in Figure~\ref{fig:max_fallback}. 
        In Table~\ref{tab:fits}, we present the knots and coefficients of such fits, so that the reader can make use of these results in a straightforward manner. 
        These quantities satisfy the following scaling relationships presented by \cite{guillochon2013} and also discussed in \cite{gafton2019}. 
        \begin{eqnarray}
            \max{\left\{\dot{M}_\text{fb}\right\}}
            &=&
            A_{5/3}\left(\frac{M_\bullet}{10^6~\text{M}_\odot}\right)^{-1/2}\left(\frac{m_*}{\text{M}_\odot}\right)^2\left(\frac{R_*}{\text{R}_\odot}\right)^{-3/2}~\text{M}_\odot~\text{yr}^{-1},\nonumber
            \\
            &&
            \\
            t_\text{peak}
            &=&
            B_{5/3}\left(\frac{M_\bullet}{10^6~\text{M}_\odot}\right)^{1/2}\left(\frac{m_*}{\text{M}_\odot}\right)^{-1}\left(\frac{R_*}{\text{R}_\odot}\right)^{3/2}~\text{yr},
            \\
            n_\infty
            &=&
            D_{5/3},
            \\
            \tau_{\text{rise},5/3}
            &=&
            K_{5/3}~\text{yr}.
        \end{eqnarray}
        The quantities $A_{5/3}$, $B_{5/3}$, and $D_{5/3}$ were introduced by \cite{guillochon2013}, where the subscript represents the adiabatic index. 
        Note that in this work we give different fits for each black hole spin case, being the superscript $``0"$, $``+"$, and $``-"$ spins $a/M_\bullet=0,~0.99$~and $-0.99$, respectively.
    \section{Coordinate transformation}
    \label{app:coordinates}
        SPHINCS uses CKS coordinates by default, thus it is necessary to bear in mind how these coordinates relate to other commonly used coordinates such as Schwarzschild, Kerr-Schild, or Boyer–Lindquist coordinates. 
        For completeness, we collect here the transformation expressions between CKS and other sets of coordinates.
        We denote the CKS coordinates as $(t,x,y,z)$. 
        These are related to the KS coordinates $(t,r,\theta,\psi)$ through the following expressions
        \begin{eqnarray}
            t
            &=&
            t_\text{KS},
            \\
            x
            &=&
            \sin\theta(r\cos\psi-a\sin\psi),
            \\
            y
            &=&
            \sin\theta(a\cos\psi+r\sin\psi),
            \\
            z
            &=&
            r\cos\theta,
        \end{eqnarray}
        and inversely,
        \begin{eqnarray}
            t_\text{KS}
            &=&
            t,
            \\
            r^2
            &=&
            \frac{1}{2}\left(x^2+y^2+z^2-a^2+\sqrt{(x^2+y^2+z^2-a^2)^2+4a^2z^2}\right),
            \nonumber
            \\
            &&
            \\
            \theta
            &=&
            \arccos\left(\frac{z}{r}\right),
            \\
            \psi
            &=&
            \arctan\left(\frac{ry-ax}{rx+ay}\right).
        \end{eqnarray}
        Note that the time coordinate is the same among these two set of coordinates.
        The velocity coordinate transformations are simply obtained by taking the respective time derivatives of these expressions, i.e.
        \begin{eqnarray}
            \dot{x}
            &=&
            \dot{\theta}\cos\theta\left(r\cos\psi-a\sin\psi\right)
            \nonumber
            \\
            &&
            +\sin\theta\left(\dot{r}\cos\psi-r\dot{\psi}\sin\psi-a\dot{\psi}\cos\psi\right)
            \\
            \dot{y}
            &=&
            \dot{\theta}\cos\theta\left(a\cos\psi+r\sin\psi\right)
            \nonumber
            \\
            &&
            +\sin\theta\left(-a\dot{\psi}\sin\psi+\dot{r}\sin\psi+r\dot{\psi}\cos\psi\right),
            \\
            \dot{z}
            &=&
            \dot{r}\cos\theta-r\dot{\theta}\sin\theta.
        \end{eqnarray}
        The Boyer-Lindquist coordinates $(t',r,\theta,\phi)$ are among the most frequently used systems to study the Kerr metric. 
        Although the $r$ and $\theta$ coordinates are identical to the Kerr-Schild coordinates, the other coordinates satisfy the following relations
        \begin{eqnarray}
            dt
            &=&
            dt' + \frac{2Mr}{\Delta}dr,
            \\
            d\psi
            &=&
            d\phi+\frac{a}{\Delta}dr,
         \end{eqnarray}
        where $\Delta=r^2-2Mr+a^2$.
\end{appendix}
\end{document}